\documentclass[final]{amsart}
\usepackage{times}
\usepackage{amsfonts,amscd,amsmath,amssymb,amsthm}

\numberwithin{equation}{section}

\newcommand{\ii}{\mathrm{i}}
\newcommand{\e}{\mathrm{e}}
\newcommand{\erfc}{\mathrm{erfc}}

\newcommand{\bw}{\mathbf{w}}

\newcommand{\diag}{\mathrm{diag}}

\newcommand{\sk}{\mathsf{k}}

\renewcommand{\Im}{{\ensuremath{\mathrm{Im\,}}}}

\DeclareSymbolFont{SY}{U}{psy}{m}{n}
\DeclareMathSymbol{\emptyset}{\mathord}{SY}{'306}
\DeclareMathSymbol{\newtimes}{\mathbin}{SY}{'264}
\DeclareMathOperator*{\Bigtimes}{\newtimes}

\DeclareMathOperator*{\slim}{s-lim}

\DeclareMathOperator{\Ran}{Ran} \DeclareMathOperator{\Ker}{Ker}
\DeclareMathOperator{\Rank}{Rank} \DeclareMathOperator{\spec}{spec}
\DeclareMathOperator{\Dom}{Dom} 

\newtheorem{theorem}{Theorem}[section]{\bf}{\it}
\newtheorem{proposition}[theorem]{Proposition}{\bf}{\it}
\newtheorem{corollary}[theorem]{Corollary}{\bf}{\it}
\newtheorem{example}[theorem]{Example}{\it}{\rm}
\newtheorem{lemma}[theorem]{Lemma}{\bf}{\it}
\newtheorem{remark}[theorem]{Remark}{\it}{\rm}
\newtheorem{definition}[theorem]{Definition}{\bf}{\it}
{\bf}{\it}

\newcommand{\R}{\mathbb{R}}

\newcommand{\C}{\mathbb{C}}

\newcommand{\N}{\mathbb{N}}

\newcommand{\1}{\mathbb{I}}

\newcommand{\fS}{\mathfrak{S}}

\newcommand{\cD}{{\mathcal D}}
\newcommand{\cE}{{\mathcal E}}

\newcommand{\cG}{{\mathcal G}}
\newcommand{\cH}{{\mathcal H}}
\newcommand{\cI}{{\mathcal I}}

\newcommand{\cK}{{\mathcal K}}
\newcommand{\cL}{{\mathcal L}}
\newcommand{\cM}{{\mathcal M}}
\newcommand{\cN}{{\mathcal N}}

\newcommand{\cP}{{\mathcal P}}
\newcommand{\cQ}{{\mathcal Q}}

\newcommand{\cW}{{\mathcal W}}

\title[Laplacians on Metric Graphs]{Laplacians on Metric Graphs: Eigenvalues, Resolvents and Semigroups}

\author[V. Kostrykin]{Vadim Kostrykin}
\address{Vadim Kostrykin\\Institut f\"{u}r Mathematik, Technische Universit\"{a}t Clausthal,
Erz-\newline stra{\ss}e 1, D-38678 Clausthal-Zellerfeld, Germany}
\email{kostrykin@math.tu-clausthal.de, kostrykin@t-online.de}

\author[R. Schrader]{Robert Schrader}
\address{Robert Schrader\\ Institut f\"{u}r Theoretische Physik\\
Freie Universit\"{a}t Berlin, Arnim\-allee 14\\ D-14195 Berlin, Germany}
\email{schrader@physik.fu-berlin.de}



\AtBeginDocument{{\noindent \textsf{\small Proceedings of the Conference on\\ Quantum Graphs and Their Applications\\
Snowbird, Utah,\\
June 18 -- June 24,  2005} \vspace{15mm}}}

\begin{document}

\begin{abstract}
The main objective of the present work is to study the negative spectrum of
(differential) Laplace operators on metric graphs as well as their
resolvents and associated heat semigroups. We prove an upper bound on the
number of negative eigenvalues and a lower bound on the spectrum of Laplace
operators. Also we provide a sufficient condition for the associated heat
semigroup to be positivity preserving.
\end{abstract}

\maketitle


\section{Introduction}

As suggested by the title, the main objective of the present work is to
study the negative spectrum of (differential) Laplace operators on metric
graphs as well as their resolvents and associated heat semigroups. Basic
notions related to these operators are summarized in Section
\ref{sec:basic} below. More complete accounts can be found in \cite{KS1},
\cite{KS8}, \cite{Kuchment:00}.

It is well known (see, e.g., \cite{D}, \cite{Ouh}) that there are deep
interrelations between pro\-perties of heat semigroups and spectral
properties of their generators. More importantly, heat semigroups generated
by Laplace-Beltrami operators on Riemannian manifolds carry a large amount
of information on the geometry of the underlying manifolds. As for metric
graphs some results in this direction can be found, e.g., in \cite{Roth:1},
\cite{Roth:2}. Nevertheless, a systematic analysis of heat semigroups for
Laplace operators on metric graphs is still missing. In this work we
perform a first step in closing this gap.

Below we will prove an upper bound on the number of negative eigenvalues
(Theorem \ref{thm:number}) and a lower bound on the spectrum of Laplace
operators (Theorem \ref{thm:3:10}). In particular, the upper bound provides
a very simple sufficient condition for the Laplace operator to be
nonnegative. The lower bound improves a recent result by Kuchment
\cite{Kuchment:00}.

Concerning resolvents and heat kernels we will establish sufficient
conditions for them to be positivity preserving. To achieve this we will
provide a closed expression for Green's function in terms of the boundary
conditions defining the corresponding Laplace operator. As a consequence
for a class of boundary conditions associated with positive maximal
isotropic subspaces (Definition \ref{def:positive} below) we prove
positivity of Green's function (see Theorems \ref{green:main} and
\ref{thm:5.1}). The proof of Theorem \ref{thm:5.1} utilizes walks on a
graph, a concept introduced in our paper \cite{KS8} as a main technical
tool for solving the inverse scattering problem on metric graphs. By
standard arguments these results imply that the associated heat semigroup is
positivity preserving.

For earlier work on heat semigroups generated by Laplace operators on metric
graphs and their application to spectral analysis we refer to \cite{AGHKH},
\cite{Angad-Guar}, \cite{Gaveau:1}, \cite{Gaveau:2}, \cite{Nicaise},
\cite{Nicaise:2}, \cite{Roth:1}, \cite{Roth:2}. Green's functions have been
studied in \cite{AGHKH}, \cite{Albeverio}, \cite{KS1}, \cite{Pokorny}.

There is a well-known approach to prove the simplicity of the lowest
eigenvalue based on the analysis of heat semigroups (see, e.g., \cite{RS}).
Although we do not pursue this in detail, merely as an illustration, for a
class of boundary conditions we prove simplicity of the lowest eigenvalue
of Laplace operators on metric graphs without internal edges (see
Proposition \ref{propo:6:4} below).

In Section \ref{sec:examples} we consider Laplace operators on metric graphs
with no internal edges, for which the heat kernel can be computed
explicitly.

\subsection*{Acknowledgements}
The authors are indebted to S.~Fulling, H.~Hofer, M.~Karowski, W.~Kirsch,
M.~Loss, and M.~Taylor for useful comments. We thank the anonymous
re\-ferees for careful reading of the manuscript and helpful suggestions.
The participation of the first author (V.K.) at the conference ``Quantum
Graphs and Their Applications'' held in the summer 2005 in Snowbird, Utah,
USA has been supported by the Deutsche Forschungsgemeinschaft and by the
National Science Foundation.

\section{Basic Structures}\label{sec:basic}

In this section we revisit the theory of Laplace operators on a metric graph
$\cG$. The material presented here is borrowed from our preceding papers
\cite{KS1} and \cite{KS8}.

A finite graph is a 4-tuple $\cG=(V,\cI,\cE,\partial)$, where $V$ is a
finite set of \emph{vertices}, $\cI$ is a finite set of \emph{internal
edges}, $\cE$ is a finite set of \emph{external edges}. Elements in
$\cI\cup\cE$ are called \emph{edges}. The map $\partial$ assigns to each
internal edge $i\in\cI$ an ordered pair of (possibly equal) vertices
$\partial(i):=\{v_1,v_2\}$ and to each external edge $e\in\cE$ a single
vertex $v$. The vertices $v_1=:\partial^-(i)$ and $v_2=:\partial^+(i)$ are
called the \emph{initial} and \emph{terminal} vertex of the internal edge
$i$, respectively. The vertex $v=\partial(e)$ is the initial vertex of the
external edge $e$. If $\partial(i)=\{v,v\}$, that is,
$\partial^-(i)=\partial^+(i)$ then $i$ is called a \emph{tadpole}. A graph
is called \emph{compact} if $\cE=\emptyset$, otherwise it is
\emph{noncompact}.

Throughout the whole work we will assume that the graph $\cG$ is connected,
that is, for any $v,v^\prime\in V$ there is an ordered sequence $\{v_1=v,
v_2,\ldots, v_n=v^\prime\}$ such that any two successive vertices in this
sequence are adjacent. In particular, this implies that any vertex of the
graph $\cG$ has nonzero degree, i.e., for any vertex there is at least one
edge with which it is incident.

We will endow the graph with the following metric structure. Any internal
edge $i\in\cI$ will be associated with an interval $[0,a_i]$ with $a_i>0$
such that the initial vertex of $i$ corresponds to $x=0$ and the terminal
one - to $x=a_i$. Any external edge $e\in\cE$ will be associated with a
semiline $[0,+\infty)$. We call the number $a_i$ the length of the internal
edge $i$. The set of lengths $\{a_i\}_{i\in\cI}$, which will also be treated
as an element of $\R^{|\cI|}$, will be denoted by $\underline{a}$. A
compact or noncompact graph $\cG$ endowed with a metric structure is called
a \emph{metric graph} $(\cG,\underline{a})$.

Given a finite graph $\cG=(V,\cI,\cE,\partial)$ with a metric structure
$\underline{a}=\{a_i\}_{i\in\cI}$ consider the Hilbert space
\begin{equation}\label{hilbert}
\cH\equiv\cH(\cE,\cI,\underline{a})=\cH_{\cE}\oplus\cH_{\cI},\qquad
\cH_{\cE}=\bigoplus_{e\in\cE}\cH_{e},\qquad
\cH_{\cI}=\bigoplus_{i\in\cI}\cH_{i},
\end{equation}
where $\cH_j=L^2(I_j)$ with
\begin{equation*}
I_j=\begin{cases} [0,a_j] & \text{if}\quad j\in\cI,\\  [0,\infty) &
\text{if}\quad j\in\cE.\end{cases}
\end{equation*}
Let $\overset{o}{I_j}$ be the interior of $I_j$, that is,
$\overset{o}{I_j}=(0,a_j)$ if $j\in\cI$ and $\overset{o}{I_j}=(0,\infty)$
if $j\in\cE$.

In the sequel the letters $x$ and $y$ will denote arbitrary elements of the
product set $\displaystyle\Bigtimes_{j\in\cE\cup\cI} I_j$.

By $\cD_j$ with $j\in\cE\cup\cI$ denote the set of all $\psi_j\in\cH_j$
such that $\psi_j(x)$ and its derivative $\psi^\prime_j(x)$ are absolutely
continuous and $\psi^{\prime\prime}_j(x)$ is square integrable. Let
$\cD_j^0$ denote the set of those elements $\psi_j\in\cD_j$ which satisfy
\begin{equation*}
\begin{matrix}
\psi_j(0)=0\\ \psi^\prime_j(0)=0
\end{matrix} \quad \text{for}\quad j\in\cE\qquad\text{and}\qquad
\begin{matrix}
\psi_j(0)=\psi_j(a_j)=0\\
\psi^\prime_j(0)=\psi^\prime_j(a_j)=0
\end{matrix}
\quad\text{for}\quad j\in\cI.
\end{equation*}
Let $\Delta^0$ be the differential operator
\begin{equation}\label{Delta:0}
\left(\Delta^0\psi\right)_j (x) = \frac{d^2}{dx^2} \psi_j(x),\qquad
j\in\cI\cup\cE
\end{equation}
with domain
\begin{equation*}
\cD^0=\bigoplus_{j\in\cE\cup\cI} \cD_j^0 \subset\cH.
\end{equation*}
It is straightforward to verify that $\Delta^0$ is a closed symmetric
operator with deficiency indices equal to $|\cE|+2|\cI|$.

We introduce an auxiliary finite-dimensional Hilbert space
\begin{equation}\label{K:def}
\cK\equiv\cK(\cE,\cI)=\cK_{\cE}\oplus\cK_{\cI}^{(-)}\oplus\cK_{\cI}^{(+)}
\end{equation}
with $\cK_{\cE}\cong\C^{|\cE|}$ and $\cK_{\cI}^{(\pm)}\cong\C^{|\cI|}$. Let
${}^d\cK$ denote the ``double'' of $\cK$, that is, ${}^d\cK=\cK\oplus\cK$.

For any $\displaystyle\psi\in\cD:=\bigoplus_{j\in\cE\cup\cI} \cD_j$ we set
\begin{equation}\label{lin1}
[\psi]:=\underline{\psi}\oplus \underline{\psi}^\prime\in{}^d\cK,
\end{equation}
with $\underline{\psi}$ and $\underline{\psi}^\prime$ defined by
\begin{equation}\label{lin1:add}
\underline{\psi} = \begin{pmatrix} \{\psi_e(0)\}_{e\in\cE} \\
                                   \{\psi_i(0)\}_{i\in\cI} \\
                                   \{\psi_i(a_i)\}_{i\in\cI} \\
                                     \end{pmatrix},\qquad
\underline{\psi}' = \begin{pmatrix} \{\psi_e'(0)\}_{e\in\cE} \\
                                   \{\psi_i'(0)\}_{i\in\cI} \\
                                   \{-\psi_i'(a_i)\}_{i\in\cI} \\
                                     \end{pmatrix}.
\end{equation}

Let $J$ be the canonical symplectic matrix on ${}^d\cK$,
\begin{equation}\label{J:canon}
J=\begin{pmatrix} 0& \1 \\ -\1 & 0
\end{pmatrix}
\end{equation}
with $\1$ being the identity operator on $\cK$. Consider the non-degenerate
Hermitian symplectic form
\begin{equation}\label{omega:canon}
\omega([\phi],[\psi]) := \langle[\phi], J[\psi]\rangle,
\end{equation}
where $\langle\cdot,\cdot\rangle$ denotes the scalar product in ${}^d
\cK\cong\C^{2(|\cE|+2|\cI|)}$.

A linear subspace $\cM$ of ${}^d\cK$ is called \emph{isotropic} if the form
$\omega$ vanishes on $\cM$ identically. An isotropic subspace is called
\emph{maximal} if it is not a proper subspace of a larger isotropic
subspace. Every maximal isotropic subspace has complex dimension equal to
$|\cE|+2|\cI|$.

Let $A$ and $B$ be linear maps of $\cK$ onto itself. By $(A,B)$ we denote
the linear map from ${}^d\cK=\cK\oplus\cK$ to $\cK$ defined by the relation
\begin{equation*}
(A,B)\; (\chi_1\oplus \chi_2) := A\, \chi_1 + B\, \chi_2,
\end{equation*}
where $\chi_1,\chi_2\in\cK$. Set
\begin{equation}\label{M:def}
\cM(A,B) := \Ker\, (A,B).
\end{equation}

\begin{theorem}\label{thm:3.1}
A subspace $\cM\subset{}^d\cK$ is maximal isotropic if and only if there
exist linear maps $A,\,B:\; \cK\rightarrow\cK$ such that $\cM=\cM(A,B)$ and
\begin{equation}\label{abcond}
\begin{split}
\mathrm{(i)}\; & \;\text{the map $(A,B):\;{}^d\cK\rightarrow\cK$ has maximal
rank equal to
$|\cE|+2|\cI|$,}\qquad\\
\mathrm{(ii)}\; &\;\text{$AB^{\dagger}$ is self-adjoint,
    $AB^{\dagger}=BA^{\dagger}$.}
\end{split}
\end{equation}
\end{theorem}

\begin{definition}\label{def:equiv}
The boundary conditions $(A,B)$ and $(A',B')$ satisfying \eqref{abcond} are
equivalent if the corresponding maximal isotropic subspaces coincide, that
is, $\cM(A,B)=\cM(A',B')$.
\end{definition}

The boundary conditions $(A,B)$ and $(A',B')$ satisfying \eqref{abcond} are
equivalent if and only if there is an invertible map $C:\,
\cK\rightarrow\cK$ such that $A'= CA$ and $B'=CB$ (see Proposition 3.6 in
\cite{KS8}).

By Lemma 3.3 in \cite{KS8}, a subspace $\cM(A,B)\subset{}^d\cK$ is maximal
isotropic if and only if
\begin{equation}\label{perp}
\cM(A,B)^{\perp} = \cM(B,-A).
\end{equation}
We mention also the equalities
\begin{equation}\label{frame}
\begin{split}
\cM(A,B)^\perp & = \bigl[\Ker\,(A,B)\bigr]^\perp  = \Ran\, (A,B)^\dagger,\\
\cM(A,B) & = \Ran(-B,A)^\dagger.
\end{split}
\end{equation}
In the terminology of symplectic geometry (see, e.g., Section 2.3 in
\cite{McDS}) the equalities \eqref{frame} have the following interpretation:
The matrix $(A,B)^\dagger$ is a (Lagrangian) \emph{frame} for the maximal
isotropic subspace $\cM(A,B)^{\perp}$ and the matrix $(-B,A)^\dagger$ is a
frame for $\cM(A,B)$.

There is an alternative parametrization of maximal isotropic subspaces of
${}^d\cK$ by unitary transformations in $\cK$ (see \cite{KS3} and
Proposition 3.6 in \cite{KS8}). A subspace $\cM(A,B)\subset {}^d\cK$ is
maximal isotropic if and only if for an arbitrary $\sk\in\R\setminus\{0\}$
the operator $A+\ii\sk B$ is invertible and
\begin{equation}\label{uuu:def}
\fS(\sk;A,B):=-(A+\ii\sk B)^{-1} (A-\ii\sk B)
\end{equation}
is unitary. Moreover, given any $\sk\in\R\setminus\{0\}$ the correspondence
between maximal isotropic subspaces $\cM\subset {}^d\cK$ and unitary
operators $\fS(\sk;A,B)\in\mathsf{U}(|\cE|+2|\cI|)$ on $\cK$ is one-to-one.
Therefore, we will use the notation $\mathfrak{S}(\sk;\cM)$ for
$\fS(\sk;A,B)$ with $\cM(A,B)=\cM$.

There is a one-to-one correspondence between all self-adjoint extensions of
$\Delta^0$ and maximal isotropic subspaces of ${}^d\cK$ (see \cite{KS1},
\cite{KS8}). In explicit terms, any self-adjoint extension of $\Delta^0$ is
the differential operator defined by \eqref{Delta:0} with domain
\begin{equation}\label{thru}
\Dom(\Delta)=\{\psi\in\cD|\; [\psi]\in\cM\},
\end{equation}
where $\cM$ is a maximal isotropic subspace of ${}^d\cK$. Conversely, any
maximal isotropic subspace $\cM$ of ${}^d\cK$ defines through \eqref{thru}
a self-adjoint operator $\Delta(\cM, \underline{a})$. If $\cI=\emptyset$,
we will simply write $\Delta(\cM)$. In the sequel we will call the operator
$\Delta(\cM, \underline{a})$ a Laplace operator on the metric graph $(\cG,
\underline{a})$. {}From the discussion above it follows immediately that
any self-adjoint Laplace operator on $\cH$ equals
$\Delta(\cM,\underline{a})$ for some maximal isotropic subspace $\cM$.
Moreover, $\Delta(\cM,\underline{a})=\Delta(\cM^\prime,\underline{a})$ if
and only if $\cM=\cM^\prime$.

{}From Theorem \ref{thm:3.1} it follows that the domain of the Laplace
operator $\Delta(\cM,\underline{a})$ consists of functions $\psi\in\cD$
satisfying the boundary conditions
\begin{equation}\label{lin2}
 A\underline{\psi}+B\underline{\psi}^{\prime}=0,
\end{equation}
with $(A,B)$ subject to \eqref{M:def} and \eqref{abcond}. Here
$\underline{\psi}$ and $\underline{\psi}^{\prime}$ are defined by
\eqref{lin1:add}.

With respect to the orthogonal decomposition $\cK =
\cK_{\cE}\oplus\cK_{\cI}^{(-)}\oplus\cK_{\cI}^{(+)}$ any element $\chi$ of
$\cK$ can be represented as a vector
\begin{equation}\label{elements}
\chi=\begin{pmatrix}\{\chi_e\}_{e\in\cE}\\ \{\chi^{(-)}_i\}_{i\in\cI}\\
\{\chi^{(+)}_i\}_{i\in\cI}\end{pmatrix}.
\end{equation}
Consider the orthogonal decomposition
\begin{equation}\label{K:ortho}
\cK = \bigoplus_{v\in V} \cL_{v}
\end{equation}
with $\cL_{v}$ the linear subspace of dimension $\deg(v)$ spanned by those
elements \eqref{elements} of $\cK$ which satisfy
\begin{equation}
\label{decomp}
\begin{split}
\chi_e=0 &\quad \text{if}\quad e\in \cE\quad\text{is not incident with the vertex}\quad v,\\
\chi^{(-)}_i=0 &\quad \text{if}\quad v\quad\text{is not an initial vertex of}\quad i\in \cI,\\
\chi^{(+)}_i=0 &\quad \text{if}\quad v\quad\text{is not a terminal vertex
of}\quad i\in \cI.
\end{split}
\end{equation}
Obviously, the subspaces $\cL_{v_1}$ and $\cL_{v_2}$ are orthogonal if
$v_1\neq v_2$.

Set ${^d}\cL_v:=\cL_v\oplus\cL_v\cong\C^{2\deg(v)}$. Obviously, each
$^d\cL_v$ inherits a symplectic structure from ${}^d\cK$ in a canonical way,
such that the orthogonal decomposition
\begin{equation*}
\bigoplus_{v\in V} {^d}\cL_v = {}^d\cK
\end{equation*}
holds.

\begin{definition}\label{propo}
Given the graph $\cG=\cG(V,\cI,\cE,\partial)$, boundary conditions $(A,B)$
satisfying \eqref{abcond} are called \emph{local on} $\cG$ if the maximal
isotropic subspace $\cM(A,B)$ of $\cK$ has an orthogonal symplectic
decomposition
\begin{equation}\label{propo:ortho}
\cM(A,B)=\bigoplus_{v\in V}\;\cM(v),
\end{equation}
with $\cM(v)$ being maximal isotropic subspaces of ${^d}\cL_v$.

Otherwise the boundary conditions are called \emph{non-local}.
\end{definition}

By Proposition 4.2 in \cite{KS8}, given the graph
$\cG=\cG(V,\cI,\cE,\partial)$, the boundary conditions $(A,B)$ satisfying
\eqref{abcond} are local on $\cG$ if and only if there is an invertible map
$C:\, \cK\rightarrow\cK$ and linear transformations $A(v)$ and $B(v)$ in
$\cL_{v}$ such that the simultaneous orthogonal decompositions
\begin{equation}\label{permut}
CA= \bigoplus_{v\in V} A(v)\quad \text{and}\quad CB= \bigoplus_{v\in V} B(v)
\end{equation}
are valid.

Given a graph $\cG=(V,\cI,\cE,\partial)$ to any vertex $v\in V$ we associate the
graph $\cG_v=(\{v\},\cI_v,\cE_v,\partial_v)$ with the following properties
\begin{itemize}
\item[(i)]{$\cI_v=\emptyset$,}
\item[(ii)]{$\partial_v(e)=v$ for all $e\in\cE_v$,}
\item[(iii)]{$|\cE_v|=\deg_{\cG}(v)$, the degree of the vertex $v$ in the graph $\cG$,}
\item[(iv)]{there is an injective map $\Psi_v:\; \cE_v\rightarrow\cE\cup\cI$ such that
$v\in\partial\circ\Psi_v(e)$ for all $e\in\cE_v$.}
\end{itemize}

Boundary conditions $(A(v), B(v))$ on each of the graphs $\cG_v$ induce
local boundary conditions $(A,B)$ on the graph $\cG$ with
\begin{equation*}
A= \bigoplus_{v\in V} A(v)\quad \text{and}\quad B= \bigoplus_{v\in V} B(v).
\end{equation*}

\begin{example}[Standard boundary conditions]\label{3:ex:3}
Given a graph $\cG$ with $|V|\geq 2$ and minimum degree not less than two,
define the boundary conditions $(A(v),B(v))$ on $\cG_v$ for every $v\in V$
by the $\deg(v)\times\deg(v)$ matrices
\begin{equation*}
\begin{aligned}
A(v)= \begin{pmatrix}
    1&-1&0&\ldots&&0&0\\
    0&1&-1&\ldots&&0 &0\\
    0&0&1&\ldots &&0 &0\\
    \vdots&\vdots&\vdots&&&\vdots&\vdots\\
    0&0&0&\ldots&&1&-1\\
    0&0&0&\ldots&&0&0
     \end{pmatrix},\quad
B(v)= \begin{pmatrix}
    0&0&0&\ldots&&0&0\\
    0&0&0&\ldots&&0&0\\
    0&0&0&\ldots&&0&0\\
    \vdots&\vdots&\vdots&&&\vdots&\vdots\\
    0&0&0&\ldots&&0&0\\
    1&1&1&\ldots&&1&1
\end{pmatrix}.
\end{aligned}
\end{equation*}
Obviously, $A(v)B(v)^\dagger=0$ and $(A(v),B(v))$ has maximal rank. Thus,
for every $v\in V$ the boundary conditions $(A(v),B(v))$ define
self-adjoint Laplace operators $\Delta(A(v),B(v))$ on $L^2(\cG_v)$. The
corresponding unitary matrices \eqref{uuu:def} are given by
\begin{equation}\label{standard:ee}
[\mathfrak{S}(\sk;A(v),B(v))]_{e,e^\prime}=
\frac{2}{\deg(v)} - \delta_{e,e^\prime}
\end{equation}
with $\delta_{e,e^\prime}$ Kronecker symbol.

The boundary conditions $(A(v),B(v))$ induce \emph{standard} local boundary
conditions $(A,B)$ on the graph $\cG$.
\end{example}

The following result is Corollary 5 in \cite{Kuchment:00}.

\begin{lemma}\label{Kuchment}
Boundary conditions $(A,B)$ satisfying \eqref{abcond} are equivalent to the
boundary conditions $(\widehat{A},\widehat{B})$ with
\begin{equation}\label{AhutBhut}
\widehat{A}= P_{\Ker\; B} + L,\qquad \widehat{B} =  P_{\Ker\; B}^\perp,
\end{equation}
where $P_{\Ker\; B}$ is the orthogonal projection in $\cK$ onto $\Ker\; B$,
$P_{\Ker\; B}^\perp:=\1-P_{\Ker\; B}$ its complementary projection, and
$L:\,\cK\rightarrow\cK$ the self-adjoint operator given by
\begin{equation*}
L=(B|_{\Ran B^\dagger})^{-1}A P_{(\Ker B)}^\perp.
\end{equation*}
\end{lemma}

Note that $\widehat{A}\widehat{B}^\dagger = L$ and $\Ker L \supset\Ker B$.

In particular, if $\Ker B=\{0\}$ such that $L=B^{-1} A$, then
$\displaystyle\mathfrak{S}(\sk;A,B) = -\frac{L+\ii\sk}{L-\ii\sk}$ (cf.\
Proposition 3.19 in \cite{KS8}).

\begin{corollary}\label{cor:Kuchment}
Assume that $\det(A-\varkappa B)\neq 0$ for some $\varkappa>0$. Then
$\mathfrak{S}(\ii\varkappa;A,B)$ is self-adjoint. If $L\leq 0$, then
$\mathfrak{S}(\ii\varkappa;A,B)$ is a contraction for all $\varkappa>0$.
\end{corollary}

\begin{proof}
The assumption $\det(A-\varkappa B)\neq 0$ combined with the fact that the
subspace $\Ker B$ reduces the operator $L$ implies that $L-\varkappa$ is
invertible. By Lemma \ref{Kuchment} we have
\begin{equation}\label{sigma:L}
\begin{split}
\mathfrak{S}(\ii\varkappa;A,B)=\mathfrak{S}(\ii\varkappa;\widehat{A},\widehat{B})
& = -P_{\Ker\; B}-P_{\Ker\; B}^\perp(L-\varkappa)^{-1}(L+\varkappa)P_{\Ker\; B}^\perp\\
& = - 2P_{\Ker\; B} - (L-\varkappa)^{-1}(L+\varkappa),
\end{split}
\end{equation}
which is self-adjoint. {}From \eqref{sigma:L} it follows that
\begin{equation*}
\|\mathfrak{S}(\ii\varkappa;A,B)\|=\max\{1,\left\|(L-\varkappa)^{-1}(L+\varkappa)\right\|\}.
\end{equation*}
If $L\leq 0$, then by the spectral theorem
$\left\|(L-\varkappa)^{-1}(L+\varkappa)\right\|\leq 1$ for all
$\varkappa>0$.
\end{proof}

\section{Eigenvalues of Laplace Operators}\label{sec:Laplace}

If $\psi=\{\psi_j\}_{j\in\cI\cup\cE}\in\cH$ is an eigenfunction of the
operator $-\Delta(A,B,\underline{a})$ corresponding to the eigenvalue
$\sk^2\in\R\setminus\{0\}$, $\Im\sk\geq 0$, then it is necessarily of the
form
\begin{equation}\label{10}
\psi_{j}(x;\sk)=\begin{cases} s_j \e^{\ii\sk x_j} &\text{for}
                                          \;j\in\cE, \\
                                  \alpha_j \e^{\ii\sk x_j}+
        \beta_j \e^{-\ii\sk x_j} & \text{for}\; j\in\cI. \end{cases}
\end{equation}
The vectors $s=\{s_e\}_{e\in\cE}\in\cK_{\cE}$,
$\alpha=\{\alpha_i\}_{i\in\cI}\in\cK_{\cI}^{(-)}$, and
$\beta=\{\beta_i\}_{i\in\cI}\in\cK_{\cI}^{(+)}$ satisfy the homogeneous
equation
\begin{equation}\label{11}
Z(\sk;A,B,\underline{a})\begin{pmatrix} s\\
                      \alpha\\
                    \beta\end{pmatrix}
                      =0
\end{equation}
with
\begin{equation}\label{Z:def}
Z(\sk;A,B,\underline{a})= A X(\sk;\underline{a})+\ii\sk B
Y(\sk;\underline{a}),
\end{equation}
where
\begin{equation}
\label{zet}
X(\sk;\underline{a})=\begin{pmatrix}\1&0&0\\
                                  0&\1&\1\\
               0&\e^{\ii\sk\underline{a}}&\e^{-\ii\sk\underline{a}}
               \end{pmatrix}\qquad\text{and}\qquad
Y(\sk;\underline{a})=\begin{pmatrix}\1&0&0\\
                                  0&\1&-\1\\
               0&-\e^{\ii\sk\underline{a}}&\e^{-\ii\sk\underline{a}}
               \end{pmatrix}.
\end{equation}
The diagonal $|\cI|\times |\cI|$ matrices $\e^{\pm \ii\sk\underline{a}}$
are given by
\begin{equation}
\label{diag} [\e^{\pm \ii\sk\underline{a}}]_{jk}=\delta_{jk}\e^{\pm \ii\sk
a_{j}}\quad
                       \text{for}\quad j,k\in\;\cI.
\end{equation}
The converse statement is also true and we have the following result.

\begin{lemma}\label{lem:3:1}
A $\psi$ given by \eqref{10} is an eigenfunction of
$-\Delta(\cM,\underline{a})$ corresponding to the eigenvalue
$\sk^2\in\R\setminus\{0\}$ if and only if \eqref{11} possesses a nontrivial
solution. The multiplicity of this eigenvalue is equal to $\dim\Ker
Z(\sk;A,B,\underline{a})$.
\end{lemma}

\begin{proof}
If $\sk^2<0$, the claim is obvious, since
$\psi\in\Dom(\Delta(\cM,\underline{a}))$. If $\sk^2>0$, any solution of
\eqref{11} satisfies $s=0$ (see the proof of Theorem 3.1 in \cite{KS1}).
Therefore, $\psi\in\Dom(\Delta(\cM,\underline{a}))$.
\end{proof}

As proven in \cite[Theorem 3.1]{KS1} the set of zeros of $\det
Z(\sk;A,B,\underline{a})$ is discrete.

Denote
\begin{equation}\label{T:def:neu}
T(\sk;\underline{a}) := \begin{pmatrix} 0 & 0 & 0 \\ 0 & 0 &
\e^{\ii\sk\underline{a}} \\ 0 & \e^{\ii\sk\underline{a}} & 0
\end{pmatrix}
\end{equation}
with respect to the orthogonal decomposition \eqref{K:def}.

\begin{theorem}\label{thm:2:6}
Assume that $\cI\neq\emptyset$. Then $\lambda=\sk^2\neq 0$ such that
$\det(A+\ii\sk B)\neq 0$ is an eigenvalue of
$-\Delta(\cM(A,B),\underline{a})$ with multiplicity $m$ if and only if $1$
is an eigenvalue of
\begin{equation*}
\mathfrak{S}(\sk;A,B) T(\sk;\underline{a})
\end{equation*}
with geometric multiplicity $m$.

If $\cI=\emptyset$, then $\lambda=-\varkappa^2<0$ is an eigenvalue of
$-\Delta(\cM(A,B))$ with multiplicity $m$ if and only if $0$ is an
eigenvalue of $A-\varkappa B$ with geometric multiplicity $m$.
\end{theorem}

\begin{remark}
In \cite{KS1} it is shown that $\det(A+\ii\sk B)\neq 0$ and $\det(A-\ii\sk
B)\neq 0$ hold for all $\sk>0$.
\end{remark}

\begin{proof}
Observe that if $\det(A+\ii\sk B)\neq 0$, then
\begin{equation}\label{verloren:3:neu}
\begin{split}
A X(\sk;\underline{a})&+\ii\sk B Y(\sk;\underline{a}) = (A+\ii\sk
B)R_+(\sk;\underline{a}) + (A-\ii\sk B)R_-(\sk;\underline{a})\\
&=(A+\ii\sk B)\big[\1+(A+\ii\sk B)^{-1}(A-\ii\sk
B)T(\sk;\underline{a})\big]R_+(\sk;\underline{a})\\
&=(A+\ii\sk B)\big[\1-\mathfrak{S}(\sk;A,B)
T(\sk;\underline{a})\big]R_+(\sk;\underline{a}),
\end{split}
\end{equation}
where
\begin{equation}\label{U:def:neu}
R_+(\sk;\underline{a}):=\frac{1}{2}[X(\sk;\underline{a})+Y(\sk;\underline{a})]
=
\begin{pmatrix} \1 & 0 & 0 \\ 0 & \1 & 0 \\
0 & 0 & \e^{-\ii\sk\underline{a}}
\end{pmatrix},
\end{equation}
\begin{equation*}
R_-(\sk;\underline{a})
:=\frac{1}{2}[X(\sk;\underline{a})-Y(\sk;\underline{a})] = \begin{pmatrix} 0
& 0 & 0 \\ 0 & 0 & \1 \\ 0 & \e^{\ii\sk\underline{a}} & 0
\end{pmatrix}.
\end{equation*}
Hence, by Lemma \ref{lem:3:1} $\begin{pmatrix}s \\ \alpha\\ \beta
\end{pmatrix}\in\cK$ is a nontrivial solution to \eqref{11} if and only if
\begin{equation}\label{12}
\big[\1-\mathfrak{S}(\sk;A,B)
T(\sk;\underline{a})\big] \begin{pmatrix} s\\
                      \alpha\\
                    \e^{-\ii\sk\underline{a}}\beta\end{pmatrix}
                      =0.
\end{equation}
If $\cI=\emptyset$, then equation \eqref{11} simplifies to $(A+\ii\sk B) s=
0$ with $s\in\cK_{\cE}$.
\end{proof}

\subsection{Positive Laplacians}\label{sec:positive:Laplacians}

Let $\cM \subset {}^d\cK$ be a maximal isotropic subspace. Let $(A,B)$ be
arbitrary boundary conditions satisfying $\cM(A,B)=\cM$. Let
$n_+(AB^\dagger)$ ($n_-(AB^\dagger)$, $n_0(AB^\dagger)$, respectively) be
the number of positive (negative, zero, respectively) eigenvalues of
$AB^\dagger$. By Sylvester's Inertia Law
\begin{equation*}
n_\pm(CAB^\dagger C^\dagger) = n_\pm(AB^\dagger),\qquad
n_0(CAB^\dagger C^\dagger) = n_0(AB^\dagger)
\end{equation*}
for any invertible $C:\, \cK\rightarrow\cK$. Since for any such $C$
\begin{equation*}
\cM(CA,CB)=\cM(A,B),
\end{equation*}
we may define
\begin{equation*}
n_\pm(\cM):=n_\pm(AB^\dagger),\qquad n_0(\cM):=n_0(AB^\dagger)
\end{equation*}
for arbitrary boundary conditions $(A,B)$ satisfying $\cM(A,B)=\cM$.

We mention several properties of numbers $n_\pm(\cM)$ and $n_0(\cM)$. First,
by \eqref{perp},
\begin{equation}\label{perpindex}
n_{\pm}(\cM^{\perp})=n_{\mp}(\cM),\qquad n_{0}(\cM^{\perp}) = n_{0}(\cM).
\end{equation}
Second, for any unitary transformation $U$ in $\cK$ one has
\begin{equation}\label{gauge2}
n_{\pm}({}^d U\cM)=n_{\pm}(\cM),\quad n_{0}({}^d U\cM)=n_{0}(\cM),
\end{equation}
where ${}^d U$ denotes the ``double'' of $U$,
\begin{equation}\label{muinv}
{}^{d}U = \begin{pmatrix}U&0\\0&U\end{pmatrix}.
\end{equation}
Finally, we have the following obvious inequality
\begin{equation}\label{n0:geq}
n_0(\cM(A,B)) \geq \max\{\dim\Ker A,\dim\Ker B\}.
\end{equation}

The numbers $n_\pm(\cM)$ and $n_0(\cM)$ admit the following equivalent
characterization in terms of the matrices $\mathfrak{S}(\sk;\cM)$.

\begin{proposition}\label{lem:3:1:propo}
Let $\cM\subset{}^d\cK$ be maximal isotropic. For any $\sk>0$ the number of
eigenvalues of $\mathfrak{S}(\sk;\cM))$ lying in the open upper (lower,
respectively) complex half-plane is equal to $n_+(\cM)$ ($n_-(\cM)$,
respectively). The number of real eigenvalues of $\mathfrak{S}(\sk;\cM)$ is
equal to $n_0(\cM)$.
\end{proposition}

\begin{proof}
Consider
\begin{equation*}
A_{\mathfrak{S}} = -\frac{1}{2}(\mathfrak{S} - \1),\qquad
B_{\mathfrak{S}}=\frac{1}{2\ii \sk}(\mathfrak{S}+\1)
\end{equation*}
with $\mathfrak{S}=\mathfrak{S}(\sk;\cM)$. Note that
$\mathfrak{S}(\sk;\cM)=\mathfrak{S}(\sk;A_{\mathfrak{S}},B_{\mathfrak{S}})$.
Observing that
\begin{equation}\label{ab:sigma:kreuz}
A_\mathfrak{S} B_\mathfrak{S}^\dagger = \frac{1}{2\sk} \Im\, \mathfrak{S}
\equiv \frac{1}{4\ii\sk}(\mathfrak{S}-\mathfrak{S}^\dagger),
\end{equation}
we obtain that
\begin{equation*}
n_\pm(\Im\mathfrak{S}(\sk;\cM)) = n_\pm(\cM),\qquad
n_0(\Im\mathfrak{S}(\sk;\cM)) = n_0(\cM)
\end{equation*}
for any $\sk>0$.
\end{proof}

Let $\cM \subset {}^d\cK$ be a maximal isotropic subspace. For arbitrary
$\varphi,\psi\in\Dom\Delta(\cM,\underline{a})$ integrating by parts we
obtain
\begin{equation}\label{sesquilinear}
\langle\varphi, -\Delta(\cM,\underline{a})\psi\rangle_{\cH} =\sum_{j\in\cE\cup\cI} \langle\varphi^\prime,\psi^\prime\rangle_{\cH_j}
+\langle[\varphi], Q[\psi]\rangle_{{}^d\cK},
\end{equation}
where $Q=\begin{pmatrix} 0 & \1 \\ 0 & 0
\end{pmatrix}$ with respect to the orthogonal
decomposition ${}^d\cK = \cK\oplus\cK$.

Using \eqref{frame} it is an elementary exercise to check that the orthogonal projection in ${}^d\cK$ onto the subspace $\cM$ is given by
\begin{equation}\label{projection}
\begin{split}
P_{\cM} & = \begin{pmatrix} -B^\dagger \\ A^\dagger
\end{pmatrix}(A A^\dagger + B B^\dagger)^{-1} (-B, A)\\ & \equiv \begin{pmatrix} B^\dagger (A A^\dagger + B
B^\dagger)^{-1} B & - B^\dagger (A A^\dagger + B B^\dagger)^{-1} A \\
-A^\dagger (A A^\dagger + B B^\dagger)^{-1} B & A^\dagger (A A^\dagger + B
B^\dagger)^{-1} A
\end{pmatrix},
\end{split}
\end{equation}
where the block matrix notation is used with respect to the orthogonal
decomposition ${}^d\cK = \cK\oplus\cK$. Since $[\varphi]$ and $[\psi]$ in \eqref{sesquilinear} belong to $\cM$, we obtain
\begin{equation*}
\langle [\varphi], Q [\psi]\rangle_{{}^d\cK} = \langle [\varphi], Q_{\cM} [\psi]\rangle_{{}^d\cK},
\end{equation*}
where $Q_{\cM}:=P_{\cM} Q P_{\cM}$. Using \eqref{projection} we get that
\begin{equation}\label{QM}
Q_{\cM} = - \begin{pmatrix} -B^\dagger \\ A^\dagger \end{pmatrix} (A
A^\dagger + B B^\dagger)^{-1} A B^\dagger (A A^\dagger + B B^\dagger)^{-1}
(-B,A).
\end{equation}
Thus, we have proven the following result.

\begin{proposition}\label{propo:sesq}
For any maximal isotropic subspace $\cM\subset{}^d\cK$
\begin{equation}\label{sesquilinear:2}
\langle\varphi, -\Delta(\cM,\underline{a})\psi\rangle_{\cH} =\sum_{j\in\cE\cup\cI} \langle\varphi^\prime,\psi^\prime\rangle_{\cH_j}
+\langle[\varphi], Q_{\cM}[\psi]\rangle_{{}^d\cK},
\end{equation}
holds for all $\varphi,\psi\in\Dom\Delta(\cM,\underline{a})$.
\end{proposition}

\begin{remark}\label{rem:Kuchment}
Observe that $Q_{\cM}$ depends on the maximal isotropic subspace $\cM$ only
and not on the particular choice of the matrices $A$ and $B$ parametrizing
$\cM$. Thus, using the parametrization of the maximal isotropic subspace
$\cM$ by matrices $(\widehat{A},\widehat{B})$ defined in \eqref{AhutBhut}
we obtain
\begin{equation}\label{Q:M:1}
Q_{\cM} = -\begin{pmatrix} -P_{\Ker B}^\perp \\ P_{\Ker B}+L \end{pmatrix}
L(\1+L^2)^{-2}\begin{pmatrix}-P_{\Ker B}^\perp, & P_{\Ker B}+L
\end{pmatrix},
\end{equation}
where we have used the equality
\begin{equation}\label{ABL}
\widehat{A} \widehat{B}^\dagger = L.
\end{equation}
Since $\Ker L\supset\Ker B$, the equality $\widehat{A}\underline{\psi} +
\widehat{B}\underline{\psi}^\prime=0$ implies that $L^2 \underline{\psi} +
L\underline{\psi}^\prime=0$. Therefore, for any
$\psi\in\Dom(\Delta(\cM,\underline{a}))$ we obtain
\begin{equation*}
\begin{split}
\begin{pmatrix}
-P_{\Ker B}^\perp, & P_{\Ker B}+L
\end{pmatrix} [\psi] & \equiv -P_{\Ker B}^\perp\underline{\psi} +
(P_{\Ker B}+L)\underline{\psi}^\prime \\ &= -(\1+L^2) P_{\Ker
B}^\perp\underline{\psi} + P_{\Ker B} \underline{\psi}^\prime.
\end{split}
\end{equation*}
Hence,
\begin{equation*}
L(\1+L^2)^{-2} \begin{pmatrix} -P_{\Ker B}^\perp, & P_{\Ker B}+L
\end{pmatrix} [\psi] = -  (\1 + L^2)^{-1} L\underline{\psi}.
\end{equation*}
Therefore, using \eqref{Q:M:1} we obtain that
\begin{equation*}
\langle [\varphi], Q_{\cM} [\psi]\rangle_{{}^d\cK} = \langle
\underline{\varphi}, \widetilde{Q}_{\cM} \underline{\psi}\rangle_{\cK},
\end{equation*}
holds for all $\varphi,\psi\in\Dom(\Delta(\cM,\underline{a}))$, where
\begin{equation*}
\widetilde{Q}_{\cM} = \begin{pmatrix} -L & 0 \\ 0 & 0 \end{pmatrix}
\end{equation*}
with the block-matrix notation with respect to the orthogonal decomposition
${}^d\cK = \cK \oplus\cK$. Together with \eqref{sesquilinear:2} this leads
to the representation for the sesquilinear form of the operator
$-\Delta(\cM,\underline{a})$ obtained previously by Kuchment in
\cite[Theorem 9]{Kuchment:00}.
\end{remark}

{}From \eqref{QM} it follows that $n_\pm(Q_{\cM})=n_\mp(\cM)$. Thus, by
Sylvester's Inertia Law, Proposition \ref{propo:sesq} immediately implies
that $-\Delta(\cM,\underline{a})\geq 0$ in the sense of quadratic forms, if
the maximal isotropic subspace $\cM\subset{}^d\cK$ satisfies the condition
$n_+(\cM)=0$.

Using variational arguments, we prove a more general result:

\begin{theorem}\label{thm:number}
The number of negative eigenvalues of $-\Delta(\cM,\underline{a})$ counting
their multiplicities is not bigger than $n_+(\cM)$. If $\cI=\emptyset$, then
$-\Delta(\cM)$ has precisely $n_+(\cM)$ negative eigenvalues.
\end{theorem}

For the proof we need the following simple lemma.

\begin{lemma}\label{lem:3:5:neu}
Assume that $(A,B)$ satisfies \eqref{abcond}. The function
$f(\varkappa):=\det(A-\varkappa B)$ has precisely $n_+(\cM(A,B))$ positive
zeroes and $n_-(\cM(A,B))$ negative zeroes (counting multiplicity). If
$\varkappa=0$ is a zero of $f$, then its multiplicity equals $\Rank
B-n_+(\cM(A,B))-n_-(\cM(A,B))\leq n_0(\cM)$.
\end{lemma}

\begin{proof}
By Lemma \ref{Kuchment} the boundary conditions $(A,B)$ are equivalent to
the boundary conditions $(\widehat{A},\widehat{B})$ defined in
\eqref{AhutBhut}. Now assume that a $\varkappa>0$ is a zero of $f$ with
multiplicity $m\geq 1$. By \eqref{ABL}, we have
\begin{equation}\label{varvar}
\det(A-\varkappa B) = c\det(\widehat{A}-\varkappa \widehat{B}) = c
\det(L-\varkappa)
\end{equation}
with a nonzero constant $c\in\C$. Thus, $\varkappa$ is an eigenvalue of $L$
with multiplicity $m$. By \eqref{ABL} it is also an eigenvalue of
$AB^\dagger$.

The case of negative eigenvalues can be considered in exactly the same way.

Observe that by \eqref{varvar} the total number of zeroes of $f(\varkappa)$
equals the dimension of $\Ran B^\dagger$, that is, the rank of $B$. Thus,
$\varkappa=0$ is a zero of $f(\varkappa)$ with multiplicity
\begin{equation*}
\begin{split}
& \Rank B-n_+(\cM(A,B))-n_-(\cM(A,B)) \\ &\qquad\leq
|\cE|+2|\cI|-n_+(\cM(A,B))-n_-(\cM(A,B))=n_0(\cM(A,B)).
\end{split}
\end{equation*}
\end{proof}

\begin{proof}[Proof of Theorem \ref{thm:number}]
Set for brevity $m=n_+(\cM)$. Assume that $-\Delta(\cM,\underline{a})$ has
at least $m+1$ negative eigenvalues. By the Min-Max Principle the $(m+1)$-st
eigenvalue is given by
\begin{equation*}
\lambda_{m+1}=\sup_{\substack{\cN\subset\cH\\ \dim\cN=m}} \inf_{\substack{\psi\in\Dom(-\Delta(\cM,\underline{a}))\\ \|\psi\|_{\cH}=1 \\ \psi\perp \cN}}
\langle\psi, -\Delta(\cM,\underline{a})\psi\rangle_{\cH}.
\end{equation*}
In particular, this implies that
\begin{equation}\label{lower:est}
\lambda_{m+1} \geq \inf_{\substack{\psi\in\Dom(-\Delta(\cM,\underline{a}))\\ \|\psi\|_{\cH}=1 \\ \psi\perp \cN}}
\langle\psi, -\Delta(\cM,\underline{a})\psi\rangle_{\cH}
\end{equation}
for any $m$-dimensional subspace $\cN$ of $\cH$.

Let $\widehat{\cN}_m\subset{}^d\cK$ be the subspace spanned by all
eigenvectors of $Q_{\cM}$ with negative eigenvalues. Since
$\Ker\,\cQ_{\cM}\supset \cM^\perp$ the inclusion $\widehat{\cN}_m\subset\cM$
holds. We claim that there is an $m$-dimensional subspace $\cN_m\subset\cH$
such that $[\psi]\in\widehat{\cN}_m$ for any $\psi\in\cN_m$. Indeed, choose
an arbitrary basis $\chi^{(1)},\ldots,\chi^{(m)}$ in $\widehat{\cN}_m$. For
every $k\in\{1,\ldots,m\}$ solve the equation
\begin{equation*}
(-\Delta(\cM,\underline{a})-1)\psi^{(k)}=0\qquad\text{with}\qquad [\psi^{(k)}]=\chi^{(k)}.
\end{equation*}
Let $\theta:\; [0,\infty]\rightarrow [0,1]$ be an arbitrary infinitely
differentiable function with $\theta(x)=1$ for all $x\in[0,1]$ and
$\theta(x)=0$ for all $x\in[2,\infty)$. Now set
$\widetilde{\psi}^{(k)}_i(x) = \psi^{(k)}_i(x)$ for all $i\in\cI$ and
\begin{equation*}
\widetilde{\psi}^{(k)}_e(x) = \psi^{(k)}_e(x) \theta(x)
\end{equation*}
for all $e\in\cE$. Obviously, $[\widetilde{\psi}^{(k)}]=[\psi^{(k)}]$ and
$\widetilde{\psi}^{(k)}\in\Dom(-\Delta(\cM,\underline{a}))$. Since
$\psi^{(k)}$ are linearly independent, we can choose $\cN_m$ as a subspace
in $\cH$ spanned by $\psi^{(1)},\ldots,\psi^{(m)}$.

Now setting $\cN=\cN_m$ in \eqref{lower:est} and using \eqref{sesquilinear:2} we obtain that
$\lambda_{m+1}\geq 0$, which is a contradiction.

The second statement of the theorem follows immediately from Lemma \ref{lem:3:5:neu}.
\end{proof}

\subsection{Eigenvalue zero}

Obviously, if $\Delta(\cM,\underline{a})\psi=0$, then $\psi$ has to be
piecewise linear,
\begin{equation}\label{0ev1}
\psi_i(x_i)={\alpha}_i+{\beta}_i x_i,\quad i\in\cI\qquad\text{and}\qquad
\psi_e(x_e)\equiv 0, \quad e\in\cE.
\end{equation}

\begin{proposition}\label{eigenwert:null}
If $n_+(\cM)=0$, then $\dim\Ker\Delta(\cM,\underline{a})\leq n_0(\cM)$.
\end{proposition}

\begin{proof}
Since $n_+(\cM)=0$, from \eqref{QM} it follows that $Q_{\cM}\geq 0$.
Therefore, if $\psi\in\Ker\Delta(\cM,\underline{a})$, then by Proposition
\ref{propo:sesq} we have $\psi^\prime=0$. Hence, $\underline{\psi}\in\Ker
A$. Therefore, by \eqref{0ev1},
\begin{equation*}
\dim\Ker\Delta(\cM,\underline{a}) \leq \dim\Ker A.
\end{equation*}
By \eqref{n0:geq} the dimension of $\Ker A$ does not exceed $n_0(\cM)$,
which proves the claim.
\end{proof}

Solutions of the Laplace equation on \emph{infinite} periodic metric graphs
have been studied recently by Kuchment and Pinchover in
\cite{Kuchment:Pinchover}. The kernel of the Laplace operator
$-\Delta(\cM,\underline{a})$ on compact graphs with standard boundary
conditions has been studied by Kurasov and Novaszek in \cite{KuNo}.

\subsection{Lower bounds on the spectrum}

Corollary 10 in \cite{Kuchment:00} in a slightly generalized form provides the following lower bound
\begin{equation*}
-\Delta(\cM,\underline{a}) \geq -4\|L_+\|(|\cE|+2|\cI|)\max\{2\|L_+\|,\max_{i\in\cI} a_i^{-1}\},
\end{equation*}
where $L_+$ is the positive part of the self-adjoint operator $L=(B|_{\Ran
B^\dagger})^{-1}A P_{(\Ker B)^\perp}$ defined in Lemma \ref{Kuchment}.
Observe that the r.h.s.\ of this estimate depends linearly on the total
degree of the graph $\cG$, that is, on the number $\displaystyle\sum_{v\in
V}\deg(v)=|\cE|+2|\cI|$. The following theorem provides a lower bound on
the spectrum of Laplace operator, which is uniform with respect to this
quantity.

\begin{theorem}\label{thm:3:10}
Assume that $\cI\neq\emptyset$. Then
\begin{equation*}
 -\Delta(\cM,\underline{a}) \geq - s(\|L_+\|)^2,
\end{equation*}
where $s(t)$ is the unique nonnegative solution of the equation
\begin{equation}
s \tanh\left(\frac{as}{2}\right)=t,\qquad a=\min_{i\in\cI} a_i.
\end{equation}
\end{theorem}

Obviously, $s(0)=0$, $s(t)$ is increasing in $t$ and satisfies the estimate
$s(t)\geq t$ for all $t\geq 0$.

\begin{remark}\label{rem:3:9}
Assume that $\cI=\emptyset$ and $n_+(\cM)>0$. Then by Theorem \ref{thm:2:6}
and Lemma \ref{Kuchment} the number $\lambda=-\|L_+\|^2$ is the smallest
eigenvalue of $-\Delta(\cM)$.
\end{remark}

\begin{proof}[Proof of Theorem \ref{thm:3:10}]
If $L_+=0$\, the claim follows from Theorem \ref{thm:number} by \eqref{ABL}.
Thus, we may assume that $L_+\neq 0$. For brevity we set
$\varkappa_0:=\|L_+\|$. Observe that by \eqref{varvar}, the number
$\varkappa_0$ is the biggest positive solution of the equation
$\det(A-\varkappa B)=0$.

For arbitrary $\varkappa>\varkappa_0$ by \eqref{sigma:L} we have
\begin{equation}\label{bigger:1}
\|\mathfrak{S}(\ii\varkappa;A,B)\| = \frac{\varkappa+\varkappa_0}{\varkappa-\varkappa_0} > 1.
\end{equation}
By Theorem \ref{thm:2:6}, if $-\varkappa_1^2<0$ is an eigenvalue of
$-\Delta(\cM(A,B),\underline{a})$, then either
\begin{equation*}
\det(A-\varkappa_1 B)=0
\end{equation*}
or
\begin{equation*}
\det[\1-\mathfrak{S}(\ii\varkappa_1;A,B)T(\ii\varkappa_1;\underline{a})]=0
\end{equation*}
holds. In the first case $\varkappa_1=\varkappa_0$. If the second case
holds, then necessarily
\begin{equation*}
\|\mathfrak{S}(\ii\varkappa_1;A,B) T(\ii\varkappa_1;\underline{a})\|\geq 1,
\end{equation*}
and, therefore,
\begin{equation*}
\|\mathfrak{S}(\ii\varkappa_1;A,B)\| \|T(\ii\varkappa_1;\underline{a})\|\geq 1.
\end{equation*}
Therefore,
\begin{equation*}
\varkappa_1 \leq \varkappa_2:=\inf\{\varkappa>0|\,
\|\mathfrak{S}(\ii\varkappa;A,B)\| \|T(\ii\varkappa;\underline{a})\|<1\}.
\end{equation*}
{}From \eqref{bigger:1} it follows that
\begin{equation*}
\|\mathfrak{S}(\ii\varkappa;A,B)\| \|T(\ii\varkappa;\underline{a})\|\leq
\frac{\varkappa+\varkappa_0}{\varkappa-\varkappa_0} \e^{-a\varkappa}
\end{equation*}
for all $\varkappa > \varkappa_0$. Therefore, $\varkappa_2$ is not bigger then the unique positive solution of
the equation
\begin{equation*}
\frac{\varkappa+\varkappa_0}{\varkappa-\varkappa_0} \e^{-a\varkappa} = 1.
\end{equation*}
It is straightforward to verify that this solution is given by
$s(\varkappa_0)$ and $s(\varkappa_0) \geq \varkappa_0$.
\end{proof}

\section{The Resolvent}\label{sec:resolvent}

In this section we will study the resolvent of the Laplace operator on a
metric graph $(\cG,\underline{a})$. In particular, we show that the
resolvent is an integral operator. The structure of the underlying Hilbert
space $\cH$ \eqref{hilbert} naturally gives rise to the following
definition of integral operators.

\begin{definition}
The operator $K$ on the Hilbert space $\cH$ is called \emph{integral
operator} if for all $j,j^\prime\in\cE\cup\cI$ there are measurable
functions $K_{j,j^\prime}(\cdot,\cdot)\, :\, I_j\times
I_{j\prime}\rightarrow \C$ with the following properties
\begin{itemize}
\item[(i)]{$K_{j,j^\prime}(x_j,\cdot)\varphi_{j^\prime}(\cdot)\in L^1(I_{j^\prime})$
for almost all $x_j\in I_j$,}
\item[(ii)]{$\psi=K\varphi$ with
\begin{equation}\label{kern}
\psi_j(x_j) = \sum_{j^\prime\in\cE\cup\cI} \int_{I_{j^\prime}}
K_{j,j^\prime}(x_j,y_{j^\prime}) \varphi_{j^\prime}(y_{j^\prime})
dy_{j^\prime}.
\end{equation}}
\end{itemize}
The $(|\cI|+|\cE|)\times(|\cI|+|\cE|)$ matrix-valued function $(x,y)\mapsto
K(x,y)$ with
\begin{equation*}
[K(x,y)]_{j,j^\prime} = K_{j,j^\prime}(x_j,y_{j^\prime})
\end{equation*}
is called the \emph{integral kernel} of the operator $K$.
\end{definition}

Below we will use the following shorthand notation for \eqref{kern}:
\begin{equation*}
\psi(x) = \int^{\cG} K(x,y) \varphi(y) dy.
\end{equation*}

\begin{lemma}\label{lem:Green}
For any maximal isotropic subspace $\cM\subset{}^d\cK$ the resolvent
\begin{equation*}
(-\Delta(\cM;\underline{a})-\sk^2)^{-1}\quad\text{for} \quad
\sk^2\in\C\setminus\mathrm{spec}(-\Delta(\cM;\underline{a}))
\end{equation*}
is the integral operator with the $(|\cI|+|\cE|)\times(|\cI|+|\cE|)$
matrix-valued integral kernel $r_{\cM}(x,y;\sk,\underline{a})$, $\Im\sk>0$,
admitting the representation
\begin{equation}\label{r:M}
\begin{split}
& r_{\cM}(x,y;\sk,\underline{a})  = r^{(0)}(x,y,\sk)\\ & - \frac{\ii}{2\sk}
\Phi(x,\sk) Z(\sk;A,B,\underline{a})^{-1} (A-\ii\sk B)
R_+(\sk;\underline{a})^{-1} \Phi(y,\sk)^T,
\end{split}
\end{equation}
where $Z(\sk;A,B,\underline{a})$ is defined in \eqref{Z:def}, the matrix
$\Phi(x,\sk)$ is given by
\begin{equation*}
\Phi(x,\sk) := \begin{pmatrix}\phi(x,\sk) & 0 & 0 \\ 0 & \phi_+(x,\sk) &
\phi_-(x,\sk)
\end{pmatrix}
\end{equation*}
with diagonal matrices $\phi(x,\sk)=\diag\{\e^{\ii\sk x_j}\}_{j\in\cE}$,
$\phi_\pm(x,\sk)=\diag\{\e^{\pm\ii\sk x_j}\}_{j\in\cI}$, and
\begin{equation*}
[r^{(0)}(x,y,\sk)]_{j,j^\prime} = \ii\delta_{j,j^\prime}
\frac{\e^{\ii\sk|x_j-y_j|}}{2\sk},\quad x_j,y_j\in I_j.
\end{equation*}
If $\det(A+\ii\sk B)\neq 0$, then
\begin{equation}\label{r:M:alternativ}
\begin{split}
& r_{\cM}(x,y;\sk,\underline{a})  = r^{(0)}(x,y,\sk)\\ & + \frac{\ii}{2\sk}
\Phi(x,\sk) R_+(\sk;\underline{a})^{-1}[\1-\mathfrak{S}(\sk;\cM)
T(\sk;\underline{a})]^{-1}\mathfrak{S}(\sk;\cM)R_+(\sk;\underline{a})^{-1}
\Phi(y,\sk)^T,
\end{split}
\end{equation}
where $R_+(\sk;\underline{a})$ is defined in \eqref{U:def:neu}. If
$\cI=\emptyset$, this representation simplifies to
\begin{equation*}
r_{\cM}(x,y,\sk) = r^{(0)}(x,y,\sk) + \frac{\ii}{2\sk} \phi(x,\sk)
\mathfrak{S}(\sk;\cM) \phi(y,\sk).
\end{equation*}
\end{lemma}

The integral kernel $r_{\cM}(x,y;\sk,\underline{a})$ is called
\emph{Green's function} or \emph{Green's matrix}.

An alternative representation for Green's function of the Laplace operator
on a graph with $\cI=\emptyset$ has been given in \cite{Albeverio}.

\begin{proof}
Since $\sk^2$ is not an eigenvalue of $-\Delta(\cM;\underline{a})$, by
Lemma \ref{lem:3:1}, the inverse in \eqref{r:M} exists. Define the operator
$M(\sk)$ as the integral operator with kernel \eqref{r:M}.

Let $\varphi\in\cH$ be arbitrary. Set $\psi=M(\sk)\varphi$. Obviously,
$\psi(x)$ is bounded and $\psi\in\cH$. To prove that
$(-\Delta(\cM;\underline{a})-\sk^2)^{-1}=M(\sk)$, it suffices to show that
\begin{itemize}
\item[(i)]{$\psi\in\Dom(\Delta(\cM;\underline{a}))$,}
\item[(ii)]{$(-\Delta(\cM;\underline{a})-\sk^2)\psi=\varphi$,}
\item[(iii)]{the symmetry relation
\begin{equation}\label{Green:sym}
r_{\cM}(y,x;\sk,\underline{a})=r_{\cM}(x,y;-\overline{\sk},\underline{a})^\dagger
\end{equation}
holds.}
\end{itemize}

\emph{Proof of (i).} Obviously, $\psi\in\cD$, where
$\displaystyle\cD=\bigoplus_{j\in\cE\cup\cI}\cD_j$ and $\cD_j$ denotes the
set of all $f_j\in\cH_j$ such that $f_j(x_j)$ and its derivative
$f_j^\prime(x_j)$ are absolutely continuous and $f_j^{\prime\prime}(x_j)$
is square integrable. Set for brevity
\begin{equation*}
G(\sk) := -Z(\sk;A,B,\underline{a})^{-1} (A-\ii\sk B)
R_+(\sk;\underline{a})^{-1}.
\end{equation*}

Assume that $\varphi_j\in\cH_j$ vanishes in a neighborhood of $x_j=0$ and,
in addition, in a neighborhood of $x_j=a_j$ if $j\in\cI$. Then
\begin{equation*}
\int_{I_j} \e^{\ii\sk|x_j-y_j|} \varphi_j(y_j) dy_j = \int_{I_j}
\e^{-\ii\sk(x_j-y_j)} \varphi_j(y_j) dy_j
\end{equation*}
holds for all sufficiently small $x_j\in I_j$ and
\begin{equation*}
\int_{I_j} \e^{\ii\sk|x_j-y_j|} \varphi_j(y_j) dy_j = \int_{I_j}
\e^{\ii\sk(x_j-y_j)} \varphi_j(y_j) dy_j
\end{equation*}
holds for all $x_j\in I_j$ sufficiently close to $a_j$ if $j\in\cI$. A
simple calculation leads to
\begin{equation*}
\underline{\psi} = \frac{\ii}{2\sk} R_+(\sk;\underline{a})^{-1}
\int^{\cG}\Phi(y,\sk)^T \varphi(y) dy + \frac{\ii}{2\sk}
X(\sk;\underline{a}) G(\sk) \int^{\cG}\Phi(y,\sk)^T \varphi(y) dy
\end{equation*}
and
\begin{equation*}
\underline{\psi}^\prime = \frac{1}{2} R_+(\sk;\underline{a})^{-1}
\int^{\cG}\Phi(y,\sk)^T \varphi(y) dy - \frac{1}{2} Y(\sk;\underline{a})
G(\sk) \int^{\cG}\Phi(y,\sk)^T \varphi(y) dy,
\end{equation*}
where $\underline{\psi}$ and $\underline{\psi}^\prime$ are defined by
\eqref{lin1:add}. Therefore,
\begin{equation*}
\begin{split}
A \underline{\psi} + B \underline{\psi}^\prime = & \frac{\ii}{2\sk}
(A-\ii\sk B)
R_+(\sk;\underline{a})^{-1} \int^{\cG}\Phi(y,\sk)^T \varphi(y) dy \\
& + \frac{\ii}{2\sk} Z(\sk; A, B, \underline{a}) G(\sk)
\int^{\cG}\Phi(y,\sk)^T \varphi(y) dy = 0.
\end{split}
\end{equation*}
Thus, we proved that $A \underline{\psi} + B \underline{\psi}^\prime = 0$
for all $\varphi$ in a dense subset of $\cH$. Therefore, $A \underline{\psi}
+ B \underline{\psi}^\prime = 0$ for all $\varphi\in\cH$, which proves the
claim (i).

\emph{Proof of (ii).} Assume that $\varphi_j\in\cH_j$ is continuous on $I_j$
for every $j\in\cI\cup\cE$. Standard arguments based on the Fourier
transform show that
\begin{equation*}
-\frac{\ii}{2\sk}\left(\frac{d^2}{dx_j^2} + \sk^2\right) \int_{I_j}
\e^{\ii\sk|x_j-y_j|} \varphi_j(y_j) dy_j = \varphi_j(x_j),\qquad x_j\in
\overset{o}{I}_j,\qquad j\in\cE\cup\cI.
\end{equation*}
where $\overset{o}{I}_j$ denotes the interior of $I_j$. This implies that
\begin{equation*}
-\frac{\ii}{2\sk}\left(\frac{d^2}{dx_j^2} + \sk^2\right) \psi_j(x_j) =
\varphi_j(x_j)
\end{equation*}
for all $x_j\in \overset{o}{I}_j$. Hence,
$(-\Delta(\cM;\underline{a})-\sk^2)\psi=\varphi$ for all $\varphi$ in a
dense subset of $\cH$. Since $M(\sk)$ is bounded, the claim follows.

\emph{Proof of (iii).} It suffices to prove \eqref{Green:sym} for the
representation \eqref{r:M:alternativ}. For the representation \eqref{r:M}
the symmetry relation  \eqref{Green:sym} will follow by continuity from
\eqref{ZZZ}.

The relation $r^{(0)}(y,x,\sk) = r^{(0)}(x,y,-\overline{\sk})^\dagger$ is
obvious. Observe that $R_+(\sk;\underline{a})^\dagger =
R_+(-\overline{\sk};\underline{a})$. Similarly,
\begin{equation*}
T(\sk;\underline{a})^\dagger =
T(-\overline{\sk};\underline{a})\qquad\text{and}\qquad \Phi(x,\sk)^\dagger
= \Phi(x,-\overline{\sk})^T\quad \text{(transpose)}.
\end{equation*}
The proof of the identity
\begin{equation*}
\mathfrak{S}(\sk;\cM)[\1-T(\sk;\underline{a})\mathfrak{S}(\sk; \cM)]^{-1} =
[\1-\mathfrak{S}(\sk; \cM) T(\sk;\underline{a})]^{-1} \mathfrak{S}(\sk;\cM).
\end{equation*}
is elementary and left to the reader. Now combining these relations we
obtain
\begin{equation*}
\begin{split}
& r_{\cM}(x,y,-\overline{\sk},\underline{a})^\dagger =  r^{(0)}(y,x,\sk)\\ &
+\frac{\ii}{2\sk}\Phi(y,\sk) R_+(\sk;\underline{a})^{-1} [\1-
\mathfrak{S}(-\overline{\sk};\cM)^\dagger T(\sk;\underline{a})]^{-1}
\mathfrak{S}(-\overline{\sk};\cM)^\dagger R_+(\sk;\underline{a})^{-1}
\Phi(x,\sk)^T.
\end{split}
\end{equation*}
Using the symmetry property
\begin{equation*}
\mathfrak{S}(-\overline{\sk};\cM)^\dagger = \mathfrak{S}(\sk;\cM),
\end{equation*}
we obtain relation \eqref{Green:sym}.

\emph{Proof of \eqref{r:M:alternativ}.} Recall that if $\det(A+\ii\sk B)\neq
0$, then
\begin{equation}\label{ZZZ}
Z(\sk; A, B, \underline{a}) = (A+\ii\sk B)[\1 - \mathfrak{S}(\sk;A,B)
T(\sk;\underline{a})] R_+(\sk;\underline{a}).
\end{equation}
Thus, \eqref{r:M:alternativ} follows from \eqref{r:M}.
\end{proof}

Observe that by Corollary \ref{cor:Kuchment} for any maximal isotropic
subspace $\cM\subset{}^d\cK$ the operator $\mathfrak{S}(\ii\varkappa;\cM)$
is self-adjoint for all $\varkappa>0$. Moreover,
$\mathfrak{S}(\ii\varkappa;\cM)$ is real if and only if the matrices
$(A,B)$ defining $\cM$ via \eqref{M:def} can be chosen real. Indeed, by
\eqref{sigma:L}, the matrix $\mathfrak{S}(\ii\varkappa;\cM)$ is real if and
only if $L$ is.

If $n_+(\cM)=0$, then again by Corollary \ref{cor:Kuchment},
$\mathfrak{S}(\ii\varkappa;\cM)$ is a self-adjoint contraction for all
$\varkappa>0$. Therefore, $\1+\mathfrak{S}(\ii\varkappa;\cM)$ is a
nonnegative operator whenever $\varkappa>0$.

\begin{definition}\label{def:order:rel}
For any square matrix $C$ we write $C\succcurlyeq 0$ (respectively, $C\succ
0$) if all entries of the matrix $C$ are nonnegative (respectively,
positive). We write $C_1\succcurlyeq C_2$ (respectively, $C_1\succ C_2$) if
$C_1-C_2\succcurlyeq 0$ (respectively, $C_1-C_2\succ 0$).
\end{definition}

\begin{definition}\label{def:positive}
The maximal isotropic subspace $\cM\subset{}^d\cK$ is called
\emph{positive}, if there is a $\varkappa_0\geq 0$ such that
$\1+\mathfrak{S}(\ii\varkappa;\cM)\succcurlyeq 0$ for all $\varkappa\geq
\varkappa_0$. It is called \emph{strictly positive}, if
$\1+\mathfrak{S}(\ii\varkappa;\cM)\succ 0$ for all $\varkappa\geq
\varkappa_0$. It is called \emph{locally strictly positive}, if the
boundary conditions defined by $\cM$ are local in the sense of Definition
\ref{propo} and $\1+\mathfrak{S}(\ii\varkappa;\cM(v))\succ 0$ for all
$\varkappa\geq \varkappa_0$ and all $v\in V$. Here $\cM(v)$ denotes the the
maximal isotropic subspace from the orthogonal decomposition
\eqref{propo:ortho}.
\end{definition}

Obviously, strictly positive maximal isotropic subspaces are positive. If
$\cM$ defines local boundary conditions on the graph $\cG$, then there are
permutation matrices $\Pi\in\mathsf{U}(|\cE|+2|\cI|)$ such that $\Pi
\mathfrak{S}(\ii\varkappa;\cM) \Pi^{-1}$ is block-diagonal with blocks
$\{\mathfrak{S}(\ii\varkappa;\cM(v))\}_{v\in V}$ (see discussion in
\cite{KS8}). Therefore, since every element of the matrix $\Pi$ is either
$1$ or $0$, if $\cM$ is locally strictly positive, then $\cM$ is positive
but not strictly positive.

\begin{example}\label{beispiel}
Consider the standard boundary conditions defined in Example \ref{3:ex:3}.
{}From \eqref{standard:ee} it follows that
\begin{equation*}
[\1+\mathfrak{S}(\sk; \cM(v))]_{e,e^\prime} = \frac{2}{\deg(v)} > 0
\end{equation*}
for all $e,e^\prime\in\cE_v$. Thus, the standard boundary conditions are
strictly positive on any graph $\cG_v$ and locally strictly positive on the
graph $\cG$.
\end{example}

\begin{theorem}\label{green:main}
Assume that the maximal isotropic subspace $\cM$ is strictly positive. Then
the Green function $r_{\cM}(x,y;\ii\varkappa,\underline{a})\succ 0$ for all
sufficiently large $\varkappa\geq 0$. Moreover, if the graph has no
internal lines ($\cI=\emptyset$), then
$r_{\cM}(x,y;\ii\varkappa,\underline{a})\succcurlyeq 0$ for all sufficiently
large $\varkappa\geq 0$ whenever the maximal isotropic subspace $\cM$ is
positive.
\end{theorem}

The case of locally strictly positive maximal isotropic subspaces will be
treated in the following section.

For the proof of Theorem \ref{green:main} we need the following lemma.

\begin{lemma}\label{lem:4.6}
Assume that $\cI\neq\emptyset$. If the maximal isotropic subspace $\cM=\cM(A,B)$ is strictly positive,
then
\begin{equation}\label{1plus}
\1+[\1 - \mathfrak{S}(\ii\varkappa;\cM) T(\ii\varkappa;\underline{a})]^{-1}
\mathfrak{S}(\ii\varkappa;\cM) \succ 0
\end{equation}
holds for all sufficiently large $\varkappa\geq 0$.
\end{lemma}

\begin{proof}
Let $\varkappa_0>0$ be the largest solution of the equation
$\det(A-\varkappa B)=0$. By Proposition 3.11 in \cite{KS8} there are
positive numbers $\varkappa_1\geq \varkappa_0$ and $C>0$ such that
\begin{equation*}
\|\mathfrak{S}(\ii\varkappa;\cM)\|\leq C
\end{equation*}
holds for all $\varkappa\geq\varkappa_1$. Observing that
$\|T(\ii\varkappa;\underline{a})\|\leq \e^{-a\varkappa}$ with $\displaystyle
a:=\min_{j\in\cI} a_j$ for all $\varkappa>0$, we obtain that
$\|\mathfrak{S}(\ii\varkappa;\cM) T(\ii\varkappa;\underline{a})\|<1/2$ for all
$\varkappa>\varkappa_2 := \max\{\varkappa_1, a^{-1}\log (2C)\}$. Therefore,
\begin{equation}\label{reihe}
\begin{split}
& \1+ [\1 - \mathfrak{S}(\ii\varkappa;\cM)
T(\ii\varkappa;\underline{a})]^{-1} \mathfrak{S}(\ii\varkappa;\cM)\\ &
\qquad = \1+\mathfrak{S}(\ii\varkappa;\cM) + \sum_{n=1}^\infty
[\mathfrak{S}(\ii\varkappa;\cM) T(\ii\varkappa;\underline{a})]^n
\mathfrak{S}(\ii\varkappa;\cM)
\end{split}
\end{equation}
converges absolutely for all $\varkappa>\varkappa_2$. Furthermore, for all
$\varkappa>\varkappa_2$ the estimate
\begin{equation*}
\begin{split}
& \left\|\sum_{n=1}^\infty [\mathfrak{S}(\ii\varkappa;\cM)
T(\ii\varkappa;\underline{a})]^n \mathfrak{S}(\ii\varkappa;\cM)\right\|\\
&\quad\leq \sum_{n=1}^\infty
\|\mathfrak{S}(\ii\varkappa;\cM)\|^{n+1}\|T(\ii\varkappa,\underline{a})\|^n\\
&\quad \leq \frac{\|\mathfrak{S}(\ii\varkappa;\cM)\|^2
\|T(\ii\varkappa;\underline{a})\|}{1 - \|\mathfrak{S}(\ii\varkappa;\cM)
T(\ii\varkappa;\underline{a})\|}
 \leq 2 C^2 \e^{-\varkappa a}
\end{split}
\end{equation*}
holds. Since all matrix elements of $\1+\mathfrak{S}(\ii\varkappa;\cM)$ are
rational functions in $\varkappa$, which are positive for sufficiently large
$\varkappa$, we obtain the claim.
\end{proof}

\begin{proof}[Proof of Theorem \ref{green:main}]
We start with the case $\cI=\emptyset$. If $e\neq e^\prime$, then
\begin{equation*}
[r_{\cM}(x,y,\ii\varkappa)]_{e,e^\prime} =
\frac{1}{2\varkappa}\e^{-\varkappa x_e}
[\mathfrak{S}(\ii\varkappa;\cM)]_{e,e^\prime} \e^{-\varkappa y_{e^\prime}}
\geq 0
\end{equation*}
for all $x_e\in I_e$ and all $y_{e^\prime}\in I_{e^\prime}$. Now consider
the case $e = e^\prime$. Noting the inequality
\begin{equation*}
\e^{-\varkappa|x_e-y_e|} \geq \e^{-\varkappa x_e} \e^{-\varkappa y_e}
\end{equation*}
we obtain that
\begin{equation*}
\begin{split}
[r_{\cM}(x,y,\ii\varkappa)]_{e,e} & \geq \frac{1}{2\varkappa} \e^{-\varkappa
x_e} \e^{-\varkappa y_e} + \frac{1}{2\varkappa} \e^{-\varkappa x_e}
[\mathfrak{S}(\ii\varkappa;\cM)]_{e,e} \e^{-\varkappa y_{e}} \\
&=\frac{1}{2\varkappa} \e^{-\varkappa x_e} \left(1+
[\mathfrak{S}(\ii\varkappa;\cM)]_{e,e}\right)\e^{-\varkappa y_{e}}\geq 0
\end{split}
\end{equation*}
holds for all sufficiently large $\varkappa$ by the assumption
$\1+S(\ii\varkappa;\cM)\succcurlyeq 0$.

Using Lemma \ref{lem:4.6} the case $\cI\neq\emptyset$ can be treated in the
same way.
\end{proof}

\section{Local Boundary Conditions: Positivity of Green's
Function}\label{sec:local:posit}

In this section we will prove an extension of Theorem \ref{green:main} to
the case of locally strictly positive maximal isotropic subspaces.

\begin{theorem}\label{thm:5.1}
Assume that $\cI\neq \emptyset$ and the graph $\cG$ has no tadpoles. If the
maximal isotropic subspace $\cM$ is locally strictly positive, then the
Green function satisfies $r_{\cM}(x,y;\ii\varkappa,\underline{a})\succ 0$
for all sufficiently large $\varkappa\geq 0$.
\end{theorem}

The proof of this theorem is more involved than that of Theorem
\ref{green:main}. Unlike the case of (globally) positive maximal isotropic
subspaces, for locally strictly positive maximal isotropic subspace the
inequality
\begin{equation*}
\1+[\1 - \mathfrak{S}(\ii\varkappa;\cM) T(\ii\varkappa;\underline{a})]^{-1}
\mathfrak{S}(\ii\varkappa;\cM) \succcurlyeq 0
\end{equation*}
in general need not hold for all large $\varkappa>0$.

Before we turn to the proof of Theorem \ref{thm:5.1} we introduce some
notion and auxiliary results.

\subsection{Walks on Graphs}

A nontrivial walk $\bw$ on the graph $\cG$ from $j\in\cE\cup\cI$ to $j^\prime\in\cE\cup\cI$ is a sequence
\begin{equation}\label{walk:def}
\{j,v_0,j_1,v_1,\ldots,j_n,v_n,j^\prime\}
\end{equation}
such that
\begin{itemize}
\item[(i)]{$j_1,\ldots,j_n\in\cI$;}
\item[(ii)]{the vertices $v_0\in V$ and $v_n\in V$ satisfy $v_0\in\partial(j)$,
$v_0\in\partial(j_1)$, $v_n\in\partial(j^\prime)$, and $v_n\in\partial(j_n)$;}
\item[(iii)]{for any $k\in\{1,\ldots,n-1\}$ the vertex $v_k\in V$ satisfies $v_k\in\partial(j_k)$ and
$v_k\in\partial(k_{k+1})$;}
\item[(iv)]{$v_k=v_{k+1}$ for some $k\in\{0,\ldots,n-1\}$ if and only if $j_k$ is a tadpole.}
\end{itemize}
If $j,j^\prime\in\cE$ this definition is equivalent to that given in \cite{KS8}.

The number $n$ is the \emph{combinatorial length} $|\bw|_{\mathrm{comb}}$
and the number
\begin{equation*}
|\bw|=\sum_{k=1}^n a_{j_k} >0
\end{equation*}
is the \emph{metric length} of the walk $\bw$.

A \emph{trivial} walk on the graph $\cG$ from $j\in\cE\cup\cI$ to
$j^\prime\in\cE\cup\cI$ is a triple $\{j,v,j^\prime\}$ such that
$v\in\partial(j)$ and $v\in\partial(j^\prime)$. Otherwise the walk is
called nontrivial. In particular, if $\partial(j)=\{v_0,v_1\}$, then
$\{j,v_0,j\}$ and $\{j,v_1,j\}$ are trivial walks, whereas
$\{j,v_0,j,v_1,j\}$ and $\{j,v_1,j,v_0,j\}$ are nontrivial walks of
combinatorial length $1$. Both the combinatorial and metric length of a
trivial walk are zero.

We will say that the walk \eqref{walk:def} leaves the edge $j$ through the
vertex $v_0$ and enters the edge $j^\prime$ through the vertex $v_n$. A
trivial walk $\{j,v,j^\prime\}$ leaves $j$ and enters $j^\prime$ through
the same vertex $v$.

A walk $\bw=\{j,v_0,j_1,v_1,\ldots,j_n,v_n,j^\prime\}$ \emph{traverses} an
internal edge $i\in\cI$ if $j_k=i$ for some $1\leq k \leq n$. It
\emph{visits} the vertex $v$ if $v_k=v$ for some $0\leq k \leq n$. The
\emph{score} $\underline{n}(\bw)$ of a walk $\bw$ is the set
$\{n_i(\bw)\}_{i\in\cI}$ with $n_i(\bw)\geq 0$ being the number of times
the walk $\bw$ traverses the internal edge $i\in\cI$.

We say that the walk is \emph{transmitted} at the vertex $v_k$ if either
$v_k=\partial(e)$ or $v_k=\partial(e^\prime)$ or $v_k\in\partial(i_k)$,
$v_k\in\partial(i_{k+1})$, and $i_k\neq i_{k+1}$. We say that a trivial
walk from $e^\prime$ to $e$ is transmitted at the vertex
$v=\partial(e)=\partial(e^\prime)$ if $e\neq e^\prime$. Otherwise the walk
is said to be \emph{reflected}. The walk is said to be
\emph{reflectionless} if it is transmitted at any vertex visited by this
walk.

Let $\chi,\chi^\prime$ be two arbitrary distinct elements of the canonical
orthonormal basis in $\cK$, that is all components of $\chi$ are zero with
the exception of one which is equal to $1$. Let $v$ be
\begin{itemize}
\item[(i)]{the initial vertex of the internal edge $j\in\cI$ if $\chi\in \cK_{\cI}^{(-)}$,}
\item[(ii)]{the terminal vertex of the internal edge $j\in\cI$ if $\chi\in \cK_{\cI}^{(+)}$,}
\item[(iii)]{the initial vertex of the external edge $j\in\cE$ if $\chi\in \cK_{\cE}$.}
\end{itemize}
Assume that $v^\prime$ is determined by the same rule from $\chi^\prime$.

Assume that the maximal isotropic subspace $\cM\subset{}^d\cK$ defines
local boundary conditions in the sense of Definition \ref{propo}. It is
straightforward to check that for any $m\in\N$ the equality
\begin{equation*}
\begin{split}
\langle\chi, \mathfrak{S}(\sk;\cM) &
(T(\sk,\underline{a})\mathfrak{S}(\sk;\cM))^m\chi^\prime\rangle  =
\sum_{\substack{\textrm{all walks}\, \bw\, \textrm{of}\\ \textrm{combinatorial length}\, m\\
\textrm{from}\, j\, \textrm{to}\, j^\prime}} \e^{\ii\sk |\bw|}\,\,
W(\sk;\bw)
\end{split}
\end{equation*}
holds, where the weight $W(\sk;\bw)$ associated with the walk
$\bw=\{j,v,j_1,v_1,\ldots,v_{m-1},$ $j_m,v^\prime,j^\prime\}$ is given by
\begin{equation*}
\begin{split}
W(\sk;\bw) & := [\mathfrak{S}(\ii\varkappa; \cM(v))]_{j,j_1} \\ & \quad\cdot
\left(\prod_{l=1}^{m-1} [\mathfrak{S}(\ii\varkappa;
\cM(v_l))]_{j_l,j_{l+1}}\right) [\mathfrak{S}(\ii\varkappa,
\cM(v^\prime))]_{j_m,j^\prime}.
\end{split}
\end{equation*}
If $m=1$ the product in the brackets has to be replaced by $1$. For $m=0$
and $v=v^\prime$ we have a similar representation
\begin{equation*}
\langle\chi, \mathfrak{S}(\sk;\cM) \chi^\prime\rangle  =
[\mathfrak{S}(\ii\varkappa; \cM(v))]_{j,j^\prime},
\end{equation*}
which corresponds to the trivial walk $\{j,v,j^\prime\}$.

Let $\cW_{j,j^\prime}^{(\sigma,\sigma^\prime)}$,
$\sigma,\sigma^\prime\in\{+,-\}$ denote the set of all walks from $j$ to
$j^\prime$ leaving $j$ through $\partial^\sigma(j)$ and entering $j^\prime$
through $\partial^{\sigma^\prime}(j^\prime)$. Observe that these sets are
disjoint.

Lemma \ref{lem:Green} and in particular equation \eqref{r:M:alternativ}
imply the following representation for Green's matrix
\begin{equation}\label{4terms:pre}
\begin{split}
[r_{\cM}(x,y,\ii\varkappa,\underline{a})]_{j,j^\prime}  = \,\, &
[r^{(0)}(x,y,\ii\varkappa)]_{j,j^\prime}\\ &
+\frac{1}{2\varkappa}\Big(\e^{-\varkappa x_j}
\sum_{\bw\in\cW_{j,j^\prime}^{(-,-)}} W(\ii\varkappa;\bw)\, \e^{-\varkappa|\bw|} \e^{-\varkappa y_{j^\prime}}\\
& + \e^{-\varkappa (a_j-x_j)} \sum_{\bw\in\cW_{j,j^\prime}^{(+,-)}}
W(\ii\varkappa;\bw)\, \e^{-\varkappa|\bw|} \e^{-\varkappa y_{j^\prime}}\\ &
+ \e^{-\varkappa x_j} \sum_{\bw\in\cW_{j,j^\prime}^{(-,+)}}
W(\ii\varkappa;\bw)\, \e^{-\varkappa|\bw|} \e^{-\varkappa
(a_{j^\prime}-y_{j^\prime})}\\ & + \e^{-\varkappa (a_j-x_j)}
\sum_{\bw\in\cW_{j,j^\prime}^{(+,+)}} W(\ii\varkappa;\bw)\,
\e^{-\varkappa|\bw|} \e^{-\varkappa (a_{j^\prime}-y_{j^\prime})}\Big),
\end{split}
\end{equation}
which holds for all sufficiently large $\varkappa>0$.

\subsection{Proof of Theorem \ref{thm:5.1}}

Below we will present a proof of Theorem \ref{thm:5.1} for the case
$j,j^\prime\in\cI$. If one or both edges are external, the proof follows
the same lines and actually is simpler than in the case considered below.

\emph{Case I:} Assume that $j=j^\prime$. Set $v:=\partial^{-}(j)$. {}From
\eqref{r:M:alternativ} and \eqref{4terms:pre} it follows that
\begin{equation*}
[r_{\cM}(x,y,\ii\varkappa,\underline{a})]_{j,j} = \e^{-2\varkappa x_j}
\left(1+ [\mathfrak{S}(\ii\varkappa; A(v), B(v))]_{j,j}+O(\e^{-2\varkappa
a_j})\right)
\end{equation*}
holds for sufficiently large $\varkappa>0$. Since $\cM$ is locally strictly
positive, we have
\begin{equation*}
[r_{\cM}(x,y,\ii\varkappa,\underline{a})]_{j,j}>0
\end{equation*}
for all large $\varkappa>0$.

\emph{Case II:} Assume now that $j\neq j^\prime$. For arbitrary
$\sigma\in\{-,+\}$ we will write
\begin{equation*}
\overline{\sigma} := \begin{cases}-, & \text{if}\quad \sigma=+,\\  +, &
\text{if}\quad \sigma=-.\end{cases}
\end{equation*}

\begin{lemma}\label{lem:5:4}
Assume that the graph $\cG$ has no tadpoles. Let $\bw\in
\cW_{j,j^\prime}^{(\sigma,\sigma^\prime)}$ be a walk with the smallest
metric length among all walks in
$\cW_{j,j^\prime}^{(\sigma,\sigma^\prime)}$. Assume that $\bw$ is not
reflectionless. Then there is a reflectionless walk $\bw^\prime\in
\cW_{j,j^\prime}^{(\overline{\sigma},\sigma^\prime)}\cup
\cW_{j,j^\prime}^{(\sigma,\overline{\sigma^\prime})}\cup
\cW_{j,j^\prime}^{(\overline{\sigma},\overline{\sigma^\prime})}$ from $j$ to
$j^\prime$ such that
\begin{equation}\label{relations:alle}
\begin{split}
\bw &= \{j,\partial^\sigma(j),\bw^\prime\}\qquad\text{if}\quad
\bw^\prime\in\cW_{j,j^\prime}^{(\overline{\sigma},\sigma^\prime)},\\
\bw &= \{\bw^\prime,\partial^{\sigma^\prime}(j^\prime), j^\prime
\}\qquad\text{if}\quad
\bw^\prime\in\cW_{j,j^\prime}^{(\sigma,\overline{\sigma^\prime})},\\
\bw &=
\{j,\partial^\sigma(j),\bw^\prime,\partial^{\sigma^\prime}(j^\prime),j^\prime\}\qquad\text{if}\quad
\bw^\prime\in\cW_{j,j^\prime}^{(\overline{\sigma},\overline{\sigma^\prime})}.
\end{split}
\end{equation}
\end{lemma}

\begin{proof}
Since $\bw\in \cW_{j,j^\prime}^{(\sigma,\sigma^\prime)}$ is a walk with the
smallest metric length among all walks in
$\cW_{j,j^\prime}^{(\sigma,\sigma^\prime)}$, it is reflected at at least
one of the vertices in $\partial(j)$ and $\partial(j^\prime)$. With this
observation the claim is obvious.
\end{proof}

\begin{lemma}\label{lem:5:3:bis}
Assume that the walk $\bw^{(\sigma,\sigma^\prime)}\in
\cW_{j,j^\prime}^{(\sigma,\sigma^\prime)}$ is reflectionless. Set
\begin{equation*}
\begin{split}
\bw^{(\overline{\sigma},\sigma^\prime)}  & :=
\{j,\partial^{\overline{\sigma}}(j),\bw^{(\sigma,\sigma^\prime)}\} \in \cW_{j,j^\prime}^{(\overline{\sigma},\sigma^\prime)},\\
\bw^{(\sigma,\overline{\sigma^\prime})}  & :=
\{\bw^{(\sigma,\sigma^\prime)},\partial^{\overline{\sigma^\prime}}(j^\prime),j^\prime\}\in \cW_{j,j^\prime}^{(\sigma,\overline{\sigma^\prime})},\\
\bw^{(\overline{\sigma},\overline{\sigma^\prime})}  & :=
\{j,\partial^{\overline{\sigma}}(j),\bw^{(\sigma,\sigma^\prime)},\partial^{\overline{\sigma^\prime}}(j^\prime),j^\prime\}\in
\cW_{j,j^\prime}^{(\overline{\sigma},\overline{\sigma^\prime})}.
\end{split}
\end{equation*}
Then
\begin{equation}\label{4terms:red}
\begin{split}
F(x_j,y_{j^\prime},\varkappa) :=\,\, & \e^{-\varkappa x_j}
W(\ii\varkappa;\bw^{(-,-)})\, \e^{-\varkappa|\bw^{(-,-)}|} \e^{-\varkappa y_{j^\prime}}\\
& + \e^{-\varkappa (a_j-x_j)} W(\ii\varkappa;\bw^{(+,-)})\,
\e^{-\varkappa|\bw^{(+,-)}|} \e^{-\varkappa y_{j^\prime}}\\ & +
\e^{-\varkappa x_j}
 W(\ii\varkappa;\bw^{(-,+)})\,
\e^{-\varkappa|\bw^{(-,+)}|} \e^{-\varkappa (a_{j^\prime}-y_{j^\prime})}\\
& + \e^{-\varkappa (a_j-x_j)} W(\ii\varkappa;\bw^{(+,+)})\,
\e^{-\varkappa|\bw^{(+,+)}|} \e^{-\varkappa (a_{j^\prime}-y_{j^\prime})}
\end{split}
\end{equation}
is positive for all sufficiently large $\varkappa>0$ and all $x_j\in I_j$,
$y_{j^\prime}\in I_{j\prime}$.
\end{lemma}

\begin{proof}
There are four different cases according to each choice of
$(\sigma,\sigma^\prime)$. It suffices to consider one case, since the other
three cases may be discussed in the same way. We pick the case
$(\sigma,\sigma^\prime)=(-,-)$, so the walk $\bw^{(-,-)}$ is assumed to be
reflectionless. By construction we have
\begin{equation*}
\begin{split}
|\bw^{(+,-)}| & = |\bw^{(-,-)}| + a_j, \\
|\bw^{(-,+)}| & = |\bw^{(-,-)}| + a_{j\prime}, \\
|\bw^{(+,+)}| & = |\bw^{(-,-)}| + a_j + a_{j\prime}.
\end{split}
\end{equation*}
Set $v:=\partial^{+}(j)$ and $v^\prime:=\partial^{+}(j^\prime)$. Obviously,
\begin{equation*}
\begin{aligned}
W(\ii\varkappa;\bw^{(+,-)}) & =W(\ii\varkappa;\bw^{(-,-)})[\mathfrak{S}(\ii\varkappa; \cM(v))]_{j,j},\\
W(\ii\varkappa;\bw^{(-,+)}) & =W(\ii\varkappa;\bw^{(-,-)})
[\mathfrak{S}(\ii\varkappa; \cM(v^\prime))]_{j^\prime,j^\prime}, \\
W(\ii\varkappa;\bw^{(+,+)}) &
=W(\ii\varkappa;\bw^{(-,-)})[\mathfrak{S}(\ii\varkappa; \cM(v))]_{j,j}
[\mathfrak{S}(\ii\varkappa; \cM(v^\prime))]_{j^\prime,j^\prime}.
\end{aligned}
\end{equation*}
The inequalities
\begin{equation*}
[\mathfrak{S}(\ii\varkappa; \cM(v))]_{j,j}>-1,\qquad
[\mathfrak{S}(\ii\varkappa; \cM(v^\prime))]_{j^\prime,j^\prime}>-1
\end{equation*}
hold for all large $\varkappa >0$ since $\cM$ is locally strictly positive.
Hence we may write
\begin{equation*}
F(x_j,y_{j^\prime},\varkappa)=\e^{-\varkappa x_j}
W(\ii\varkappa;\bw^{(-,-)})\, \e^{-\varkappa|\bw|}\e^{-\varkappa
x_{j^\prime}} (1+a(\varkappa))(1+b(\varkappa))
\end{equation*}
where
\begin{equation*}
\begin{aligned}
-1<a(\varkappa)&=\e^{-2\varkappa(a_j-x_j)}[\mathfrak{S}(\ii\varkappa; \cM(v))]_{j,j},\\
-1<b(\varkappa)&=\e^{-2\varkappa(a_{j^\prime} -x_{j^\prime})}
[\mathfrak{S}(\ii\varkappa; \cM(v^\prime))]_{j^\prime,j^\prime}
\end{aligned}
\end{equation*}
holds for all large $\varkappa$. Now $W(\ii\varkappa;\bw^{(-,-)})>0$ since
$\bw^{(-,-)}$ is reflectionless. Therefore $F(x_j,y_{j^\prime},\varkappa)$
is positive for all large $\varkappa>0$.
\end{proof}

Now we are in the position to complete the proof of Theorem \ref{thm:5.1}.

Assume that $\bw$ is a shortest walk in
$\cW_{j,j^\prime}^{(\sigma,\sigma^\prime)}$ for some
$\sigma,\sigma^\prime\in\{+,-\}$. By Lemma \ref{lem:5:4} this walk is
either reflectionless or there is a reflectionless walk $\bw^\prime\in
\cW_{j,j^\prime}^{(\overline{\sigma},\sigma^\prime)}\cup
\cW_{j,j^\prime}^{(\sigma,\overline{\sigma^\prime})}\cup
\cW_{j,j^\prime}^{(\overline{\sigma},\overline{\sigma^\prime})}$ such that
$\bw$ can be obtained from $\bw^\prime$ by one of relations
\eqref{relations:alle}.

Applying Lemma \ref{lem:5:3:bis} to the walk $\bw$ in the first case and to
the walk $\bw^\prime$ in the second case we obtain that the corresponding
contribution to the sum in \eqref{4terms:pre} is positive. Hence, the
leading term on the r.h.s.\ of \eqref{4terms:pre} is positive for all
sufficiently large $\varkappa>0$.

\section{Positivity Preserving Heat Semigroups}\label{sec:semigroups}

Since $-\Delta(\cM;\underline{a})$ is bounded from below, the heat
semigroup $\exp\{t \Delta(\cM,\underline{a})\}$ defined by the spectral
theorem is a bounded operator.

\begin{lemma}\label{lem:semigroup}
The heat semigroup $\exp\{t \Delta(\cM,\underline{a})\}$ associated with
the Laplace operator $-\Delta(\cM;\underline{a})$ is the integral operator
with $(|\cI|+|\cE|)\times(|\cI|+|\cE|)$ matrix-valued integral kernel
$p_t(x,y;\cM,\underline{a})$. The integral kernel is bounded for every
$t>0$, that is, for any $t>0$ there is a constant $C_t>0$ such that the
bound
\begin{equation*}
|[p_t(x,y;\cM,\underline{a})]_{j,j^\prime}| \leq C_t
\end{equation*}
holds for all $\displaystyle x,y\in\Bigtimes_{j\in\cE\cup\cI} I_j$ and all
$j,j^\prime\in\cE\cup\cI$. Moreover, it is infinitely differentiable for all
$\displaystyle x,y\in\Bigtimes_{j\in\cE\cup\cI} \overset{o}{I_j}$.
\end{lemma}

Recall that $\overset{o}{I_j}$ denote the interior of $I_j$.

The integral kernel $p_t(x,y;\cM,\underline{a})$ is called the \emph{heat
kernel} associated with the Laplace operator $\Delta(\cM,\underline{a})$.
If $\cI=\emptyset$ we will write $p_t(x,y;\cM)$.

For any $p\in[1,\infty]$ we set
\begin{equation*}
L^p(\cG):=\{\psi=\{\psi_j\}_{j\in\cE\cup\cI}\,|\,\psi_j\in L^p(I_j)\}.
\end{equation*}

\begin{proof}[Proof of Lemma \ref{lem:semigroup}]
Choose $\varkappa>0$ so large that
$-\varkappa^2<\inf\spec(-\Delta(\cM,\underline{a}))$. Then
\begin{equation*}
\exp\{t \Delta(\cM,\underline{a})\} =
(\Delta(\cM,\underline{a})-\varkappa^2)^{-1}
(\Delta(\cM,\underline{a})-\varkappa^2) \exp\{t \Delta(\cM,\underline{a})\}.
\end{equation*}
By the spectral theorem $(\Delta(\cM,\underline{a})-\varkappa^2) \exp\{t
\Delta(\cM,\underline{a})\}$ is bounded as a map from $L^2(\cG)$ to itself.
By Lemma \ref{lem:Green} the resolvent
$(-\Delta(\cM;\underline{a})-z)^{-1}$ maps $L^2(\cG)$ into $L^\infty(\cG)$.
By a general theory of Carleman operators, it follows that the heat
semigroup $\exp\{t \Delta(\cM,\underline{a})\}$ is the integral operator
with essentially bounded integral kernel.

To establish smoothness, we note that by a similar argument
\begin{equation}\label{bla:bla}
(-\Delta_{\cM,\underline{a}}+\varkappa^2)^n\exp\{t\Delta_{\cM,\underline{a}}\}
(-\Delta_{\cM,\underline{a}}+\varkappa^2)^n
\end{equation}
is a bounded map from $L^2(\cG)$ into $L^{\infty}(\cG)$ for all $n\in\N$ and
all $\varkappa>0$. Thus, \eqref{bla:bla} is an integral operator with
essentially bounded integral kernel. This implies that
\begin{equation*}
\left(-\frac{d^2}{dx_j^2}+ \varkappa^2\right)^n
\left(-\frac{d^2}{dy_{j^\prime}^2}+ \varkappa^2\right)^n
[p_t(x,y;\cM,\underline{a})]_{j,j^\prime}
\end{equation*}
is bounded for almost all $x_j\in\overset{o}{I_j}$ and
$y_{j^\prime}\in\overset{o}{I_{j^\prime}}$.
\end{proof}

The self-adjointness of $\Delta(\cM,\underline{a})$ implies the following
symmetry relation of the heat kernel
\begin{equation}\label{heat:sym}
p_t(y,x;\cM,\underline{a}) = p_t(x,y;\cM,\underline{a})^\dagger.
\end{equation}

For special graphs and special boundary conditions (see \cite{AGHKH},
\cite{Angad-Guar}, \cite{Gaveau:1}, \cite{Gaveau:2} as well as Section
\ref{sec:examples} below) the integral kernel of the heat semigroup can be
computed explicitly.

\begin{definition}\label{def:6:2}
$\psi\equiv\{\psi_j\}_{j\in\cI\cup\cE}\in\cH$ is called \emph{nonnegative},
in symbols $\psi\geq 0$, if $\psi_j(x)\geq 0$ for Lebesgue almost all $x\in
I_j$ for any $j\in\cI\cup\cE$. The semigroup $\exp\{t
\Delta(\cM,\underline{a})\}$ is called \emph{positivity preserving} if
$\exp\{t \Delta(\cM,\underline{a})\} \psi\geq 0$ holds for all nonnegative
$\psi\in\cH$.
\end{definition}

\begin{theorem}\label{3:theo:1}
(i) If $\cM$ is strictly positive, then the heat semigroup
$\e^{t\Delta(\cM,\underline{a})}$ associated with the Laplace operator
$\Delta(\cM,\underline{a})$ is positivity preserving for all $t>0$. If $\cI
= \emptyset$ it suffices to assume that $\cM$ is positive.

(ii) If $\cM$ is locally strictly positive and the graph $\cG$ has no
tadpoles, then the heat semigroup $\e^{t\Delta(\cM,\underline{a})}$
associated with the Laplace operator $\Delta(\cM,\underline{a})$ is
positivity preserving for all $t>0$.
\end{theorem}

\begin{proof}[Proof of Theorem \ref{3:theo:1}]
Using the operator-valued Euler formula \cite{D}, \cite{Ouh}
\begin{equation*}
\slim_{n\rightarrow \infty}
 \left(-\frac{t}{n}\Delta(\cM,\underline{a})+1\right)^{-n} \equiv \slim_{n\rightarrow
 \infty}\left(\frac{n}{t}\right)^n
 \left(-\Delta(\cM,\underline{a})+\frac{n}{t}\right)^{-n}=
\e^{t\Delta(\cM,\underline{a})}.
\end{equation*}
the claim (i) follows from Theorem \ref{green:main} and the claim (ii) from
Theorem \ref{thm:5.1}.
\end{proof}

We close this section with a simple applications of results of the last two
sections to negative spectrum of Laplace operators.

\begin{proposition}\label{propo:6:4}
Assume that $\cI=\emptyset$. If $\cM$ is strictly positive, then the
smallest eigenvalue $\lambda\leq 0$ of $-\Delta(\cM)$ is simple.
\end{proposition}

\begin{proof}
We call $\psi\equiv\{\psi_j\}_{j\in\cE}$ positive, in symbols $\psi>0$, if
$\psi_j(x_j)>0$ for Lebesgue almost all $x_j\in I_j$ for all $j\in\cE$.
Observe that by Lemma \ref{lem:Green} and Theorem \ref{green:main} the
resolvent $(-\Delta(\cM)+\varkappa^2)^{-1}$ is positivity improving for all
$\varkappa>\sqrt{-\lambda}$, that is,
\begin{equation*}
(-\Delta(\cM)+\varkappa^2)^{-1}\psi >0
\end{equation*}
holds for all $\psi\geq 0$. Applying Theorem XIII.44 in \cite{RS} we obtain
the claim.
\end{proof}

\section{Some Examples}\label{sec:examples}

Throughout this section we assume that the connected graph $\cG$ has no
internal edges, $\cG=(\{v\},\emptyset,\cE,\partial)$ with
$\deg(v)=|\cE|\geq 1$.

For standard boundary conditions (see Example \ref{3:ex:3}) the heat kernel
has been calculated in \cite{Gaveau:2}:
\begin{equation}\label{heat:kernel:standard}
[p_t(x,y;\cM)]_{e,e^\prime}= \begin{cases} \displaystyle
g_t(x_e-y_e)-\frac{\deg(v)-2}{\deg(v)}
g_t(x_e+y_e), & e=e^\prime,\\
\displaystyle \frac{2}{\deg(v)}g_t(x_e+y_{e^\prime}), & e\neq e^\prime,
\end{cases}
\end{equation}
where
\begin{equation*}
g_t(x)=\frac{1}{\sqrt{4\pi t}} \exp\{-x^2/4t\}.
\end{equation*}
By Theorem \ref{3:theo:1} the heat semigroup associated with the heat
kernel $p_t(x,y)$ is positivity preserving (see Example \ref{beispiel}).
Alternatively, positivity of the heat kernel can be deduced directly from
\eqref{heat:kernel:standard} using the inequality $g_t(x_e+y_e)\leq
g_t(x_e-y_e)$ for all $x_e,y_e>0$.

Below we will derive an explicit representation for the heat kernel for a
class of boundary conditions which to the best of our knowledge has not been
treated before. We start with recalling the following well-known result
(see, e.g., \cite{BoFu}, \cite{CJ}).

\begin{lemma}\label{6:lem:1}
For $t>0$ the function $h_t(s,\lambda)$, defined in terms of the
complementary error function
\begin{equation*}
\erfc(w)=\frac{2}{\sqrt{\pi}}\int_w^\infty \e^{-u^2}du
\end{equation*}
as
\begin{equation}\label{htfinal}
\begin{aligned}
h_t(s,\lambda) &=-\lambda \sqrt{4\pi t}\, g_t(s)
\exp\left\{\left(\frac{s}{2\sqrt{t}}-\lambda\sqrt{t}\right)^2\right\}
\erfc\left(\frac{s}{2\sqrt{t}}-\lambda\sqrt{t}\right)\\
&=-\lambda \frac{g_t(s)}{g_t(s-2\lambda t)}
\erfc\left(\frac{s}{2\sqrt{t}}-\lambda\sqrt{t}\right),
\end{aligned}
\end{equation}
satisfies the relations
\begin{equation}\label{hrobin}
\begin{aligned}
\partial_t h_t(s;\lambda)-\partial_s^2 h_t(s;\lambda)&=0,\\
\lambda(2g_t(s)- h_t(s;\lambda))-\partial_s h_t(s;\lambda)&=0.\\
\end{aligned}
\end{equation}
\end{lemma}

{}From the well-known asymptotic expansion of the error function
\cite[Equation 7.2.14]{AS} we obtain the asymptotics
\begin{equation}\label{chiasymp}
h_t(s;\lambda) = - \frac{4 \lambda t}{s}g_t(s)(1+O(t)),
\end{equation}
which holds for small $t>0$ and fixed $s>0$.

Assume that the maximal isotropic subspace $\cM\subset{}^d\cK$ is defined
by the self-adjoint $A=H$ and $B=\1$ via relation \eqref{M:def}. Then
\begin{equation*}
\mathfrak{S}(\sk;\cM) = -\frac{H-\ii\sk}{H+\ii\sk}
\end{equation*}
such that
\begin{equation*}
\1+\mathfrak{S}(\ii\varkappa;\cM) = 2\varkappa(\varkappa - H)^{-1}.
\end{equation*}
Observe that the maximal isotropic subspace $\cM(H,\1)$ is positive if
either $H\succcurlyeq 0$ or all off-diagonal matrix elements of $H$ are
positive. Indeed, for sufficiently large $\varkappa$ we have the absolutely
convergent expansion
\begin{equation*}
\1+\mathfrak{S}(\ii\varkappa;\cM) = 2\sum_{n=0}^\infty
\left(\frac{H}{\varkappa}\right)^n.
\end{equation*}
If $H\succcurlyeq 0$ this sum is nonnegative. If all off-diagonal matrix
elements of $H$ are positive, then $H\succ - c\1$ for some $c\geq 0$. Thus,
\begin{equation*}
\1+\varkappa^{-1} H \succ 0
\end{equation*}
holds for all $\varkappa>c$. This implies that
$\1+\mathfrak{S}(\ii\varkappa;\cM)\succ 0$ for all sufficiently large
$\varkappa>0$.

For any $e,e^\prime\in\cE$ consider
\begin{equation*}
f_t^{(e,e^\prime)}(s):= [h_t(s,H)]_{e,e^\prime},
\end{equation*}
where the matrix-valued function $h_t(s,H)$ is defined by \eqref{htfinal}
via the spectral theorem.

\begin{theorem}\label{6:theo:1}
Assume that the maximal isotropic subspace $\cM\subset{}^d\cK$ is defined
by the self-adjoint $A=H$ and $B=\1$ via relation \eqref{M:def}. The heat
kernel $p_t(x,y;\cM)$ of $-\Delta(\cM)$ is given by
\begin{equation}\label{heatsvertex}
[p_t(x,y;\cM)]_{e,e^\prime} = g_t(x_e-y_e)\,\delta_{e,e^\prime} +
g_t(x_e+y_e)\,\delta_{e,e^\prime} - f_t^{(e,e^\prime)}(x_e+y_{e^\prime}).
\end{equation}
For small $t>0$ and fixed $x,y$ the following asymptotics holds
\begin{equation}\label{tsmall}
\begin{split}
[p_t(x,y;\cM)]_{e,e^\prime} & =  g_t(x_e-y_e)\,\delta_{e,e^\prime} +
g_t(x_e+y_e)\,\delta_{e,e^\prime}\\ & \quad + \frac{4 t
H_{e,e^\prime}}{(x_e+y_{e^\prime})}g_t(x_e+y_{e^\prime})(1+O(t)).
\end{split}
\end{equation}
\end{theorem}

\begin{proof}
Obviously, the integral kernel \eqref{heatsvertex} satisfies the symmetry
relation \eqref{heat:sym}. Set for brevity
\begin{equation}\label{brevity}
\cP^{(e,e^\prime)}_t(x_e,y_{e^\prime}) := g_t(x_e-y_e)\,\delta_{e,e^\prime}
+ g_t(x_e+y_e)\,\delta_{e,e^\prime}-f_t^{(e,e^\prime)}(x_e+y_{e^\prime}).
\end{equation}
By Lemma \ref{6:lem:1} for any $e,e^\prime\in\cE$ the function
$\cP^{(e,e^\prime)}_t(x_e,y_{e^\prime})$ solves the heat conduction equation
\begin{equation*}
\partial_t \cP^{(e,e^\prime)}_t(x_e,y_{e^\prime}) = \partial_{x_e}^2
\cP^{(e,e^\prime)}_t(x_e,y_{e^\prime}).
\end{equation*}
Therefore, we need only to check that \eqref{brevity} satisfies the boundary
conditions \eqref{lin2} with $A=H$ and $B=\1$.

{}From \eqref{hrobin} we get
\begin{equation*}
\partial_{s} h_t(s;H) + H h_t(s;H) = 2 g_t(s) H.
\end{equation*}
Therefore,
\begin{equation}\label{neu:ref}
\partial_{y_{e^\prime}} f^{(e,e^\prime)}_t(y_{e^\prime}) +
\sum_{e^{\prime\prime}\in\cE} H_{e,e^{\prime\prime}}
f^{(e^{\prime\prime},e^\prime)}_t(y_{e^\prime})|_{x_e=0} = 2
g_t(y_{e^\prime}) H_{e,e^\prime}.
\end{equation}
Equation \eqref{brevity} implies that
\begin{equation*}
\cP^{(e,e^\prime)}_t(x_e,y_{e^\prime})|_{x_e=0} = 2 g_t(y_e)
\delta_{e,e^\prime} - f_t^{(e,e^\prime)}(y_{e^\prime})
\end{equation*}
and
\begin{equation*}
\partial_{x_e}\cP^{(e,e^\prime)}_t(x_e,y_{e^\prime})|_{x_e=0} =
- \partial_{y_{e^\prime}} f_t^{(e,e^\prime)}(y_{e^\prime}).
\end{equation*}
{}From \eqref{neu:ref} it follows that
\begin{equation*}
\begin{split}
& \sum_{e^{\prime\prime}\in\cE} H_{e,e^{\prime\prime}}
\cP^{(e^{\prime\prime},e^\prime)}_t(x_e,y_{e^\prime})|_{x_e=0} +
\partial_{x_e}\cP^{(e,e^\prime)}_t(x_e,y_{e^\prime})|_{x_e=0} \\
& \quad = 2 g_t(y_{e^\prime}) H_{e,e^\prime}  -
\sum_{e^{\prime\prime}\in\cE} H_{e,e^{\prime\prime}}
f_t^{(e^{\prime\prime},e^\prime)}(y_{e^\prime}) - \partial_{y_{e^\prime}}
f_t^{(e,e^\prime)}(y_{e^\prime}) = 0
\end{split}
\end{equation*}
holds for all $y_{e^\prime}>0$. This proves the equality
\eqref{heatsvertex}. The asymptotics \eqref{tsmall} follows from
\eqref{chiasymp}.
\end{proof}

In particular, the asymptotics \eqref{tsmall} implies that for small $t>0$
the heat kernel is nonnegative whenever $H\succcurlyeq 0$.


\end{document}